\documentclass[12pt]{article}

\usepackage{newtxtext,newtxmath}

\usepackage{graphicx}

\usepackage[letterpaper,margin=1in]{geometry}

\renewenvironment{abstract}
	{\quotation}
	{\endquotation}

\date{}

\makeatletter
\renewcommand{\fnum@figure}{\textbf{Figure \thefigure}}
\renewcommand{\fnum@table}{\textbf{Table \thetable}}
\makeatother

\usepackage{scicite}

\usepackage{url}

\def\scititle{Global Science Sustains U.S. Innovation}
\title{\bfseries \boldmath \scititle}

\author{
	Christopher~Esposito$^{1\ast}$\and
	\small$^{1}$UCLA Anderson School of Management, Los Angeles, CA 90095, USA.\and
	\small$^\ast$Corresponding author. Email: christopher.esposito@anderson.ucla.edu
}

\begin{document}

\maketitle

\begin{abstract} \bfseries \boldmath
Like physical products, new technologies are developed using globally sourced inputs. Yet while the supply chains behind physical goods are well understood, we know far less about the international supply chain of scientific knowledge that powers U.S. innovation, or how vulnerable it may be to disruption. Here, I uncover this supply chain by tracing multi-generational citation paths connecting NSF-funded research to downstream patents, and stress-test it by simulating barriers to scientific knowledge flows across the U.S. border. The U.S. knowledge supply chain extends globally, and frictions impeding the movement of ideas across the U.S. border reduce its connectivity, extend its length, and lower innovation productivity. These impacts extend to technology areas deemed critical to national priorities by U.S. Congress, including Semiconductors, Quantum Science, and AI.
\end{abstract}

\noindent
Does U.S. innovation rely on a global supply chain of knowledge? Anecdotes like the development of CRISPR technology---initiated by foundational research in the U.S., advanced in Europe, and patented in the U.S.---suggest that it might \cite{altschul_basic_1990,tatusov_genomic_1997,benson_genbank_2013,mojica_biological_2000,jansen_identification_2002,mojica_intervening_2005,barrangou_2007,zhang_crispr-cas_2014,doudna_methods_2019}. Innovation theory also suggests that it could: both evolutionary and classical theories describe new technologies as products of accumulated knowledge \cite{romer_endogenous_1990,mokyr_lever_1990,weitzman1998,fleming_technology_2001}. As the complexity of technology has risen over time, U.S.-based inventors have come to source an ever-expanding breadth of knowledge \cite{jones_burden_2009,wagner_2005,nagar_erc_2024,wang_contribution_2024}. These developments raise the possibility that the supply chains of scientific knowledge used by the U.S. to produce new technologies may extend globally, crossing international borders multiple times before contributing to domestic innovation.

Yet empirical evidence is lacking. Save for research on the propensity for U.S.-based patents to cite science produced in China or funded by the European Research Council, there have been no attempts to systematically measure the dependence of U.S. innovation on foreign science \cite{wang_contribution_2024,nagar_erc_2024}. Part of this lapse is explained by a historical lack of urgency. Until recently, low barriers to international knowledge transfer and the United States' uncontested position as the leading technological superpower obviated the need to identify the structure of this supply chain or to test its resilience.

The unraveling of the international liberal order and the rise of China as a scientific and technological powerhouse has changed that calculus \cite{wu_chinas_2024}. Increasingly, governments treat science and technology as instruments of national security rather than as global public goods \cite{chatterji_how_2025}. This logic has been compounded by a transactional turn in foreign policy, which has strained relations even between traditional allies \cite{lind_press_2025}. Together, these forces have given rise to policies and practices that impede the movement of scientific knowledge across borders. In the United States, grants involving foreign collaborators have been canceled, export restrictions on scientific tools and datasets have been imposed, and visa regimes have been tightened \cite{wu_chinas_2024,dhs_2025,nih_foreign_subawards_2025}; meanwhile, longstanding allies and technological competitors alike are responding by redirecting scientific talent and investment away from the U.S. \cite{choose_europe_2025,caut_2025}. Interventions restricting the movement of scientists across borders carry particular risk for the diffusion of scientific knowledge, because much scientific knowledge is tacit and difficult to transmit without face-to-face communication \cite{polanyi_tacit_1966,jaffe_geographic_1993,balsmeier_science_localizes_2025,balsmeier_isolating_2023,lin_remote_2023}. These global turns raise two questions: does U.S. innovation depend on foreign scientific contributions, and what would happen to the quantity and efficiency of U.S. innovation if access to foreign science were curtailed?

Think of a new invention as the endpoint of a trajectory of knowledge development \cite{ahmadpoor_dual_2017}. While the invention itself may build upon applied science, that applied science may build upon upstream foundational research. Each stage of a knowledge trajectory is carried out in a location with advantageous factor endowments---for example, CRISPR's foundational work was completed in the U.S., where scale in research activities supported investments in genome cataloging \cite{benson_genbank_2013}; many of its conceptual developments were made in Spain, France, and the Netherlands, where scientists had experience studying halophile bacteria and manipulating yogurt cultures for longevity \cite{mojica_intervening_2005, barrangou_2007}; and its translation into patented technologies occurred in the U.S., where researchers and investors in the metropolitan areas centered on San Francisco and Boston contributed applied and translational expertise \cite{doudna_methods_2019, zhang_crispr-cas_2014}. By partitioning research tasks across localities, global knowledge supply chains expand innovation possibilities and increase its productivity \cite{samila_venture_2010,samila_venture_2011}. However, they also require the international transfer of scientific knowledge. Crucially, the barriers to sharing scientific knowledge across borders are now rising.

When the cross-border barriers rise, knowledge trajectories evolve in new directions. Less able to cross national borders or source data and tools from foreign colleagues, scientists may form new collaborations, diffuse their findings to different audiences, or learn from different peers. Such frictions alter the direction of research, the location where it is carried out, the likelihood that a given trajectory will culminate in a technological invention, and the quantity of research and development that is required to convert foundational research into new technologies.

\subsection*{Uncovering the Knowledge Supply Chain}

While citations do not capture all knowledge flows, they provide a means of tracing the knowledge supply chain, connecting foundational science to applied science and on to patents. The starting point for this analysis are 772,584 USPTO patents that cite scientific research \cite{marx_reliance_2020} and over 240 million scientific publications indexed in OpenAlex. Using the full patent-to-paper citation graph, I trace trajectories forward from instances of foundational research---papers acknowledging NSF funding \cite{lin_sciscinet_2023}---through successive layers of citing papers until reaching the nearest downstream patent (Fig.~\ref{fig:fig1}A). In total, 360,419 NSF-funded papers connect along the citation graph to downstream patents. I extract the shortest of these paths to identify most intellectually-direct routes through which NSF-funded research connects to patented inventions.

I label each step along each of these shortest paths by its citation distance to the nearest downstream patent, counting back from the patent at $Distance = 0$ \cite{ahmadpoor_dual_2017}. The median shortest path connecting an NSF-funded paper to a patent is 3 citation steps long, placing it at a ``Distance to Nearest Patent'' location of $-3$ (Fig.~\ref{fig:fig1}B), which corresponds to an average time lag of 8 years between the publication of the NSF-funded paper and the application of the patent.

To characterize the geography of these paths, I assign each paper to the country of affiliation of its last author, and each patent to the country of residence of one of its inventors chosen at random (see SM for analyses using all authors and inventors on bylines). Where appropriate, countries are aggregated to supranational regions, such as Europe and Africa.

Figure~\ref{fig:fig1}C shows the share of U.S.-produced papers and patents at each step along these trajectories, broken out by path length. The U.S. share traces a distinct U-shape. Most paths originate in the U.S., where 83.1\% of NSF-supported start papers are produced. The share then falls sharply toward the patent boundary, with 56.2\% of intermediate papers produced \textit{outside} the U.S., before rebounding at the endpoint, where 61.5\% of downstream patents are produced by U.S. inventors. This pattern holds across path lengths: U.S.-based scientists disproportionately initiate these trajectories and U.S.-based inventors disproportionately capture their endpoints, while foreign researchers contribute the majority of the science in between.

Figure~\ref{fig:fig1}D illustrates the same paths using a flow diagram, revealing how production shifts across countries and supranational regions as paths advance from origin to patent. Only the median-length paths (length 3) are shown in the figure, but patterns for the other length paths are similar. While most of the paths originate and terminate in the U.S., they frequently circulate through international regions along the way, passing through Europe, the U.K., China, East Asia, and Canada before returning to the U.S. for patenting. Among the ten non-U.S. countries and regions, only Israel exhibits a net positive inflow of science used for patenting, reflecting its well-documented technology commercialization ecosystem \cite{trajtenberg2001}.

\subsection*{Long-run impacts of knowledge flow restrictions}

How would restrictions on cross-border knowledge flows affect U.S. innovation? Each citation link crossing the U.S. border represents the movement of scientific ideas into or out of the U.S. knowledge ecosystem; frictions on cross-border flows would render some of these citations less probable, effectively removing edges from the citation graph. I model this directly by asking how the shortest paths in the citation network would reconfigure if citations crossing the U.S. border were removed. Deleted citations that fall outside existing shortest paths leave those trajectories intact, while those that fall on a shortest path force it to reconnect to the nearest patent along the shortest remaining route. Where no such route exists, the path would be lost entirely.

I simulate 11 scenarios of increasingly severe cross-border knowledge flow barriers by probabilistically deleting citations that cross the U.S. border, ranging from the status quo (0\% deleted) to complete U.S. scientific autarky (100\% deleted), with intermediate scenarios at 10\% increments. After each round of deletions, paths reconfigure along the shortest remaining routes. Figure~\ref{fig:fig1}A illustrates the logic: removing the citation linking Paper 5 to Paper 2 forces the path to reroute through Papers 4 and 1, terminating at Patent A.

To predict the impacts of border restrictions on innovation, Figure~\ref{fig:fig2}A shows \textit{path capture} in and outside the U.S.---measured as the number of paths terminating at U.S. and non-U.S. patents---across all 11 scenarios. The values are normalized to their values under the status quo, at which 223,847 paths are captured in the U.S. and 136,572 are captured outside the U.S. (see Fig.~\ref{fig:s_allregions_capture_longrun}). As border frictions rise, capture changes in both regions: very light frictions (10\%) slightly increase path capture in the U.S., as paths that would have terminated outside the U.S. are re-routed to U.S. endpoints, while such light frictions reduce path capture elsewhere. More intensive border frictions (50\%) reduce path capture in the U.S. by 11\% and outside the U.S. by 7\%. The most intensive frictions---full autarky, with a 100\% citation deletion probability---reduces U.S. capture by 16\% and non-U.S. capture by 64\%.

Notably, for all but the extreme case of U.S. autarky, the decline in U.S.-captured paths exceeds the decline in non-U.S.-captured paths. This occurs in part because U.S.-captured paths cross the U.S. border more frequently under the status quo: on average 0.48 times, compared to 0.40 times for non-U.S.-captured paths (Fig.~\ref{fig:s_border_crossings}). Since the probability of disruption compounds with each border-crossing citation, paths that cross more frequently face greater risk. Under full autarky, however, the pattern reverses: non-U.S. capture collapses because these paths cannot leave the U.S. to reach foreign patents in the first place.

In addition to these effects on U.S. and non-U.S. path capture, Figure~\ref{fig:s_allregions_capture_longrun} shows path capture for disaggregated countries and global regions. Under the status quo, 45,919 paths are captured in Europe, 12,887 in the U.K., 9,954 in China, 21,454 in East Asia (Japan, South Korea, and Taiwan), and 15,061 in Israel. Of these, Israel is most sensitive to U.S. border frictions: its captured paths decline by 83\% under full autarky, consistent with its unusually high dependence on cross-border knowledge flows documented in Figure~\ref{fig:s_allregions_border_crossings}.

In addition to reducing path capture, restrictions on international knowledge flows can reduce the innovation productivity of countries and supranational regions, as more research effort needs to be expended to complete a path connecting an NSF-funded paper to a downstream patent. At the country or supranational level, research productivity may decline through two core mechanisms.

The first mechanism is path elongation. As border frictions rise, disrupted trajectories often reconnect to patents along longer routes, increasing the research effort required to sustain each connection. Fig.~\ref{fig:s_path_length} shows that the model predicts that moving from the status quo to autarky results in an increase in mean path length from 2.98 to 3.63 for U.S.-captured paths and from 3.01 to 4.97 for non-U.S.-captured paths. The divergence is driven by paths that originate close to the paper-patent boundary: among paths starting at $Distance = -1$, autarky increases length by 30.8\% for U.S.-ending paths but 98.0\% for non-U.S.-ending paths, whereas paths starting at $Distance = -5$ lengthen by similar amounts regardless of endpoint (Fig.~\ref{fig:s_path_length_by_d}). This asymmetry reflects the greater difficulty outside the U.S. in translating applied scientific knowledge into patents.

The second mechanism is the on-shoring of research activity: as cross-border flows are restricted, the U.S. becomes more reliant on domestically produced science. Figure~\ref{fig:s_intermediary_steps_US} shows that moving from the status quo to autarky results in the share of U.S.-produced intermediary papers that connect to U.S. patents to rise from 54.7\% to 100\%, confirming that, under complete scientific isolation, the U.S. would have to supply all of the science underpinning its own innovations.

Combining these two mechanisms, I measure U.S. innovation productivity under each border friction scenario $s$ as the number of U.S.-captured paths per U.S.-produced intermediate paper on those paths:

\begin{equation}
\Pi_{s} = \frac{\text{U.S. Captured Paths}_s}{\text{U.S. Produced Intermediate Papers on U.S. Captured Paths}_s}
\label{eq:productivity}
\end{equation}

\noindent Figure~\ref{fig:fig2}B plots $\Pi_{s}$ across all scenarios, normalized to 1 under the status quo. Productivity declines steadily as border frictions rise: a 50\% deletion rate reduces productivity by 9\%, and full autarky reduces it by 50\%. These results indicate that the U.S. would have to substantially increase its domestic scientific investment if it sought to maintain its current levels of innovative output under frictions to knowledge transfer across its border.

Figure~\ref{fig:fig2}C shows the effects of knowledge flow restrictions on U.S. path capture and innovation productivity across the 11 technology areas designated as critical to national interests by the CHIPS and Science Act of 2022 \cite{chips_science_act_2022}, plus an aggregate grouping of all 11 areas. Patents were assigned to critical technology areas by using an LLM to match patent CPC subclass labels to NSF-provided keyword dictionaries for each technology area (see SM).

Border frictions reduce U.S. path capture across most critical technology areas, with full autarky producing a 20\% decline in the aggregate grouping. The productivity effects are more severe: autarky reduces innovation productivity across every critical technology field, by 48\% in the aggregate grouping, and by more than 50\% in Advanced Manufacturing, Advanced Materials, Artificial Intelligence, Communications and Wireless, Disaster Resilience, Energy Technology, and Quantum Science.

Together, these results establish that U.S. innovation is deeply reliant on foreign science, and that restrictions to cross-border knowledge flows are projected to reduce both the quantity and productive efficiency of U.S. patenting, with substantial effects concentrated in the technology areas Congress has identified as critical to national interests. These long-run estimates, however, speak only to paths already realized in the citation record. An equally important question concerns the short-run impacts of knowledge flow restrictions---the effects on the paths already under development.

\subsection*{Short-run impacts of knowledge flow restrictions}

The long-run simulations described above operate on paths already realized as patents. But restrictions on cross-border knowledge flows would also affect the paths currently under development. These \textit{outstanding paths} are NSF-supported papers that have not yet linked to a downstream patent but are expected to do so. As of 2021, the final year of the patent-to-paper citations data \cite{marx_reliance_2020}, there are 194,452 NSF-funded papers that have not yet linked to a patent.

To predict their outcomes, I draw on the \textit{realized paths}---the full set of NSF-funded papers that have already linked to a downstream patent---as an empirical baseline. I first down-sample outstanding paths using a hazard model to drop the paths unlikely to ever realize, based on their age (measured in years) relative to the empirical age-at-realization distribution of realized paths. This reduces the outstanding paths from 194,452 to 100,899. I then assign each surviving path to a $(\mathrm{StartingD}, \mathrm{CurrentD}, \mathrm{CurrentRegion})$ cell via a probabilistic draw conditioned on each path's current and projected age-at-realization, preserving relationships observed in realized paths. Finally, I use realized paths to compute transition probabilities from each cell to each possible outcome (U.S. capture, non-U.S. capture, or stalling) under each border restriction scenario. Full details are provided in the SM.

Figure~\ref{fig:fig_short_run}A shows outstanding path capture by the U.S. and non-U.S. regions across citation deletion rate scenarios, measured relative to the status quo. Moving from the status quo to autarky is projected to result in a decline from 61,407 to 39,118 outstanding paths captured in the U.S. (a 39\% decline) and from 39,239 to 15,215 outstanding paths captured outside the U.S. (a 61\% decline). These impacts are large in part because 51\% of outstanding paths are currently being advanced outside the U.S. (Fig.~\ref{fig:s_outstanding_path_assignement_regions}): many paths that would return to the U.S. for patenting under the status quo are instead stranded abroad when cross-border flows are restricted.

The economic impact of the decline in U.S. outstanding path capture can be assessed against NSF's total investment in basic research. The NSF has spent roughly \$150 billion since 2000, spread across 499,763 highly-cited NSF-funded papers, implying approximately \$300,142 per paper. Moving from the status quo to autarky is predicted to result in 22,289 U.S.-captured paths becoming disconnected from their downstream patents (Fig.~\ref{fig:s_capture_absolute_shortrun}), implying that \$6.7 billion in NSF-supported capital no longer connect to U.S. patents under autarky.

With regard to innovation productivity, Figure~\ref{fig:fig_short_run}B shows that moving from the status quo to autarky results in U.S. innovation productivity for outstanding paths to decline by 46\%, indicating that maintaining current levels of innovative output would require a substantial increase in domestic science funding.

As with the long-run impacts, the short-run impacts of border restrictions also impact U.S. innovation in congressionally-defined critical technology areas. Figure~\ref{fig:fig_short_run}C shows that knowledge flow restrictions reduce outstanding path capture and innovation productivity across all critical technologies. Under autarky, U.S. path capture declines by 49\% in the aggregate grouping, and innovation productivity falls by more than 50\% across each of the 11 individual categories.

Finally, I link downstream patents to their corporate assignees \cite{arora_discern_2023} to show how restrictions affect path capture at the 20 U.S. corporations most reliant on NSF-stimulated knowledge paths (Fig.~\ref{fig:s_firms_path_capture}). Autarky reduces capture at 19 of these 20 firms, with declines spanning software (61\% at Alphabet, 52\% at Microsoft), hardware (36\% at Intel, 64\% at Micron), energy (63\% at Halliburton, 51\% at ExxonMobil), and aerospace (57\% at Lockheed Martin, 49\% at Boeing). These results demonstrate the costs that protracted border restrictions would impose on the innovative outcomes of powerful stakeholders.

\subsection*{Discussion}

This study traced the global supply chain of scientific knowledge that sustains U.S. innovation, revealing a system that is deeply international and structurally vulnerable. The majority of intermediary papers linking NSF-funded research to downstream patents are produced outside the U.S., yet the downstream patents themselves are overwhelmingly U.S.-produced, reflecting the U.S.'s disproportionate capacity to commercialize globally-circulating knowledge. Restrictions on cross-border knowledge flows disrupt this system through two core mechanisms: lengthening the paths that connect basic science to patented invention, and trapping promising trajectories outside U.S. borders before they can be commercialized. These effects are not confined to any single domain, but are evident across the full range of technology areas designated as critical to national priorities by U.S. Congress, such as Semiconductors, Quantum Science, and Artificial Intelligence.

More broadly, these findings reveal that countries are bound together by invisible supply chains of knowledge that underpin their innovation. The knowledge that eventually becomes a U.S. patent may have passed through laboratories in Europe, Asia, Canada, or other parts of the world, each contributing a link in a chain of technological development whose full extent is rarely visible to the scientists working within them, let alone the policymakers who enact policies that shape their evolution. As governments increasingly view science and technology as competitive assets and erect barriers to their movement, they risk disrupting these supply chains, reducing their efficiency or severing them completely. The present analysis offers a framework for making these supply chains visible, and for quantifying what is at stake if they are broken.


\begin{figure}
\centering
\begin{minipage}[t]{0.32\textwidth}
\centering
\textbf{A}\\
\includegraphics[width=\textwidth]{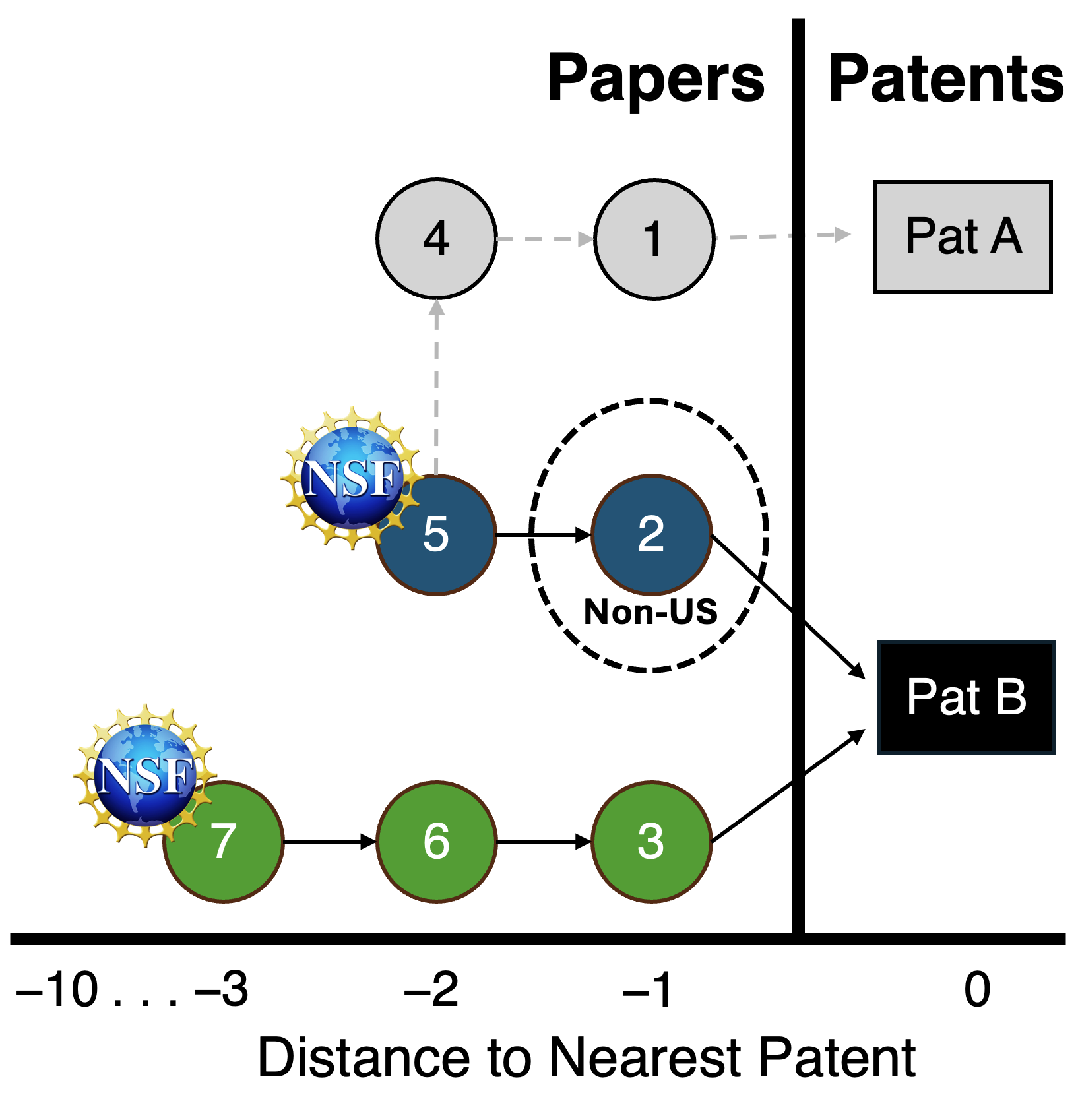}
\end{minipage}\hfill
\begin{minipage}[t]{0.32\textwidth}
\centering
\textbf{B}\\
\includegraphics[width=\textwidth]{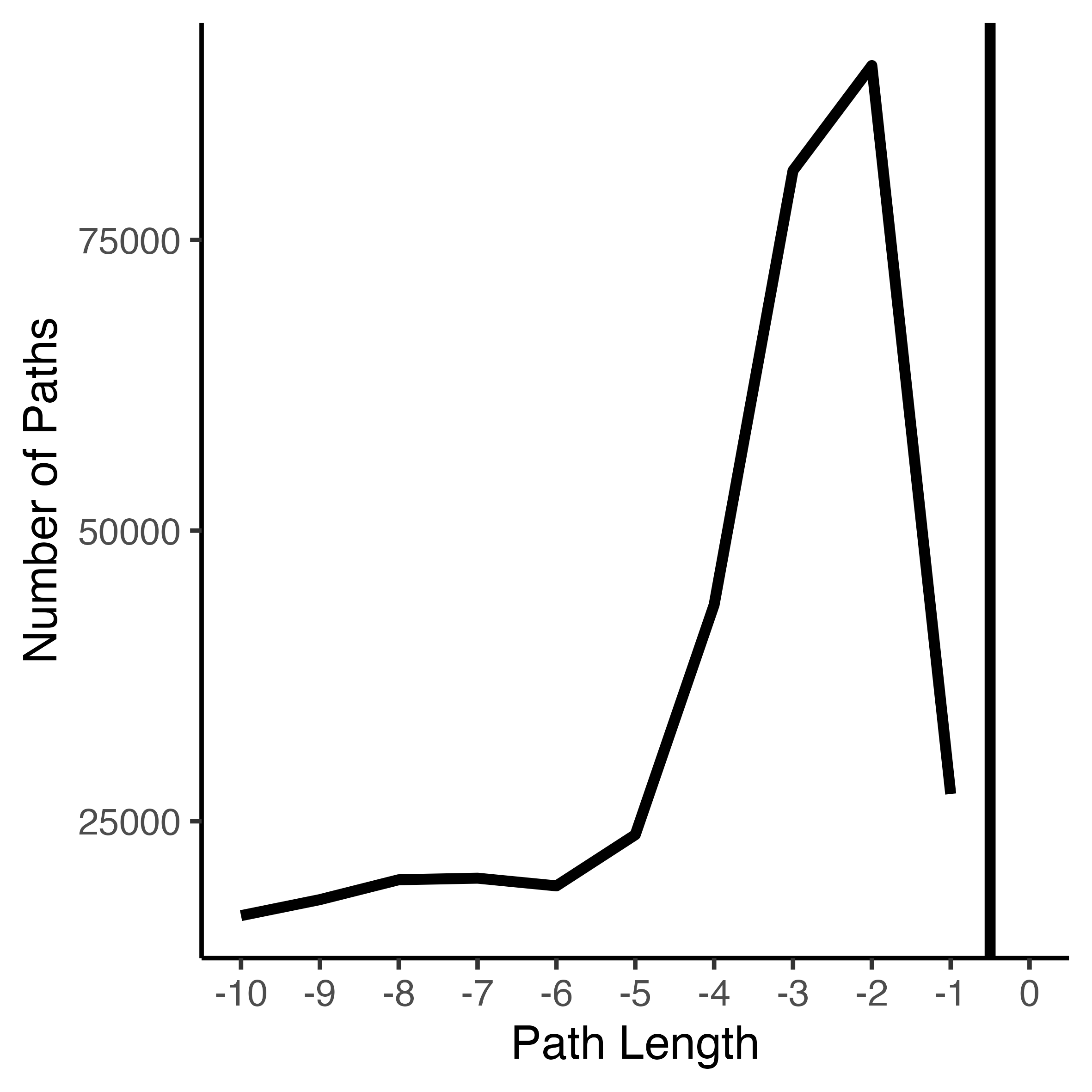}
\end{minipage}\hfill
\begin{minipage}[t]{0.32\textwidth}
\centering
\textbf{C}\\
\includegraphics[width=\textwidth]{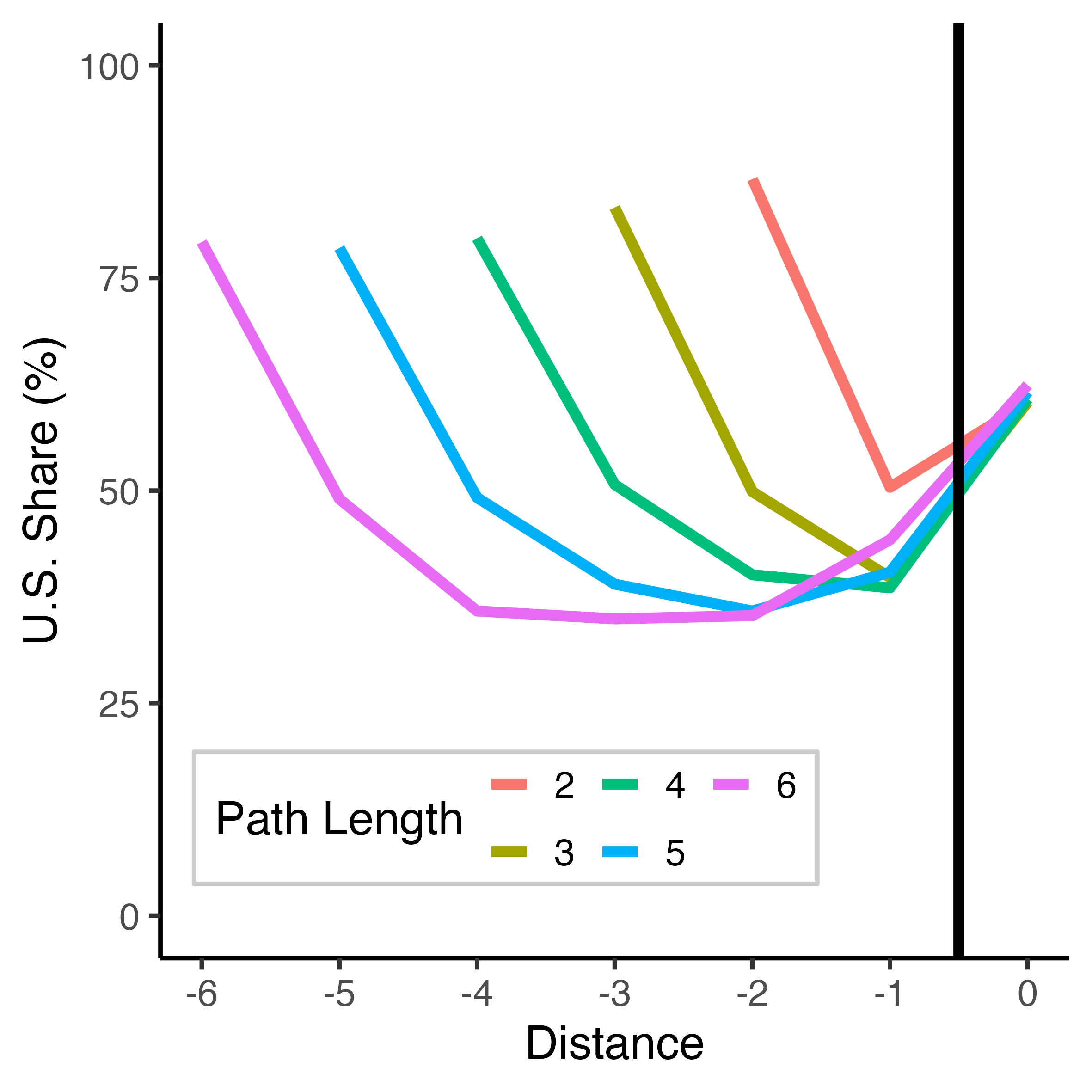}
\end{minipage}

\vspace{1em}

\begin{minipage}[t]{\textwidth}
\centering
\textbf{D}\\
\includegraphics[width=\textwidth]{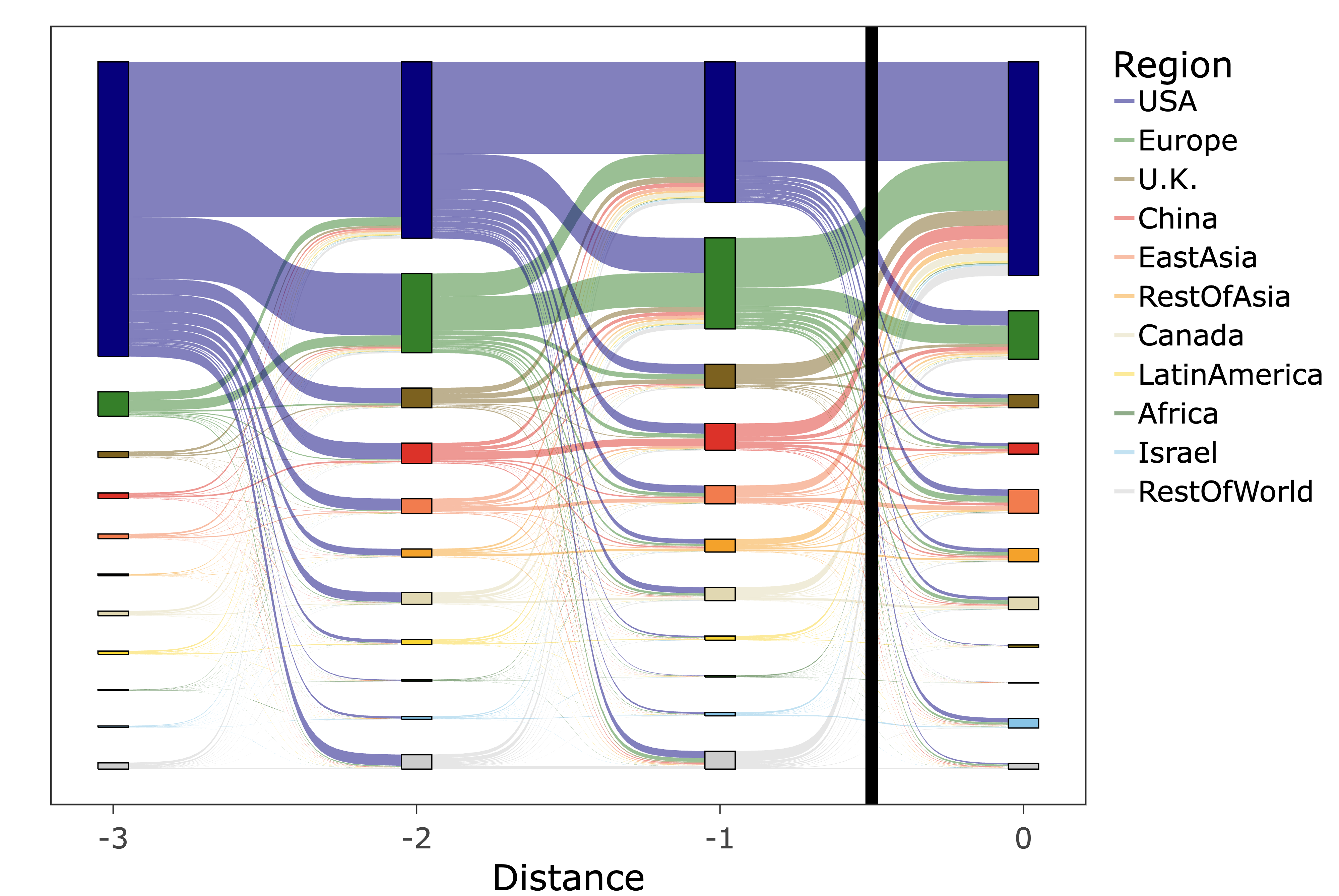}
\end{minipage}

\caption{\linespread{1.0}\selectfont\textbf{Global structure of science--technology linkages.} (\textbf{A}) Schematic of how shortest citation paths linking NSF-funded research to patents are extracted from the full citation network, and how they reconfigure under counterfactual restrictions to cross-border knowledge flow. When a cross-border citation is probabilistically removed, affected paths reroute along the shortest remaining route to the nearest patent (e.g., the grey route to Patent A). (\textbf{B}) Distribution of NSF-funded papers by citation distance to the nearest downstream patent. The median path is 3 steps long, corresponding to an average 8-year lag between initiating paper publication and patent application. (\textbf{C}) Share of U.S.-produced documents at each citation distance, by path length. The U-shaped relationship shows that the U.S. share is highest at path origins, falls sharply in intermediate steps, and rebounds at the patent endpoints. (\textbf{D}) Flow structure of length-3 citation paths showing how shortest paths circulate globally before patenting, frequently passing through Europe, the U.K., China, East Asia, and Canada.}

\label{fig:fig1}
\end{figure}

\begin{figure}
\centering
\begin{minipage}[t]{0.4\textwidth}
\centering
\textbf{A}\\
\includegraphics[width=\textwidth]{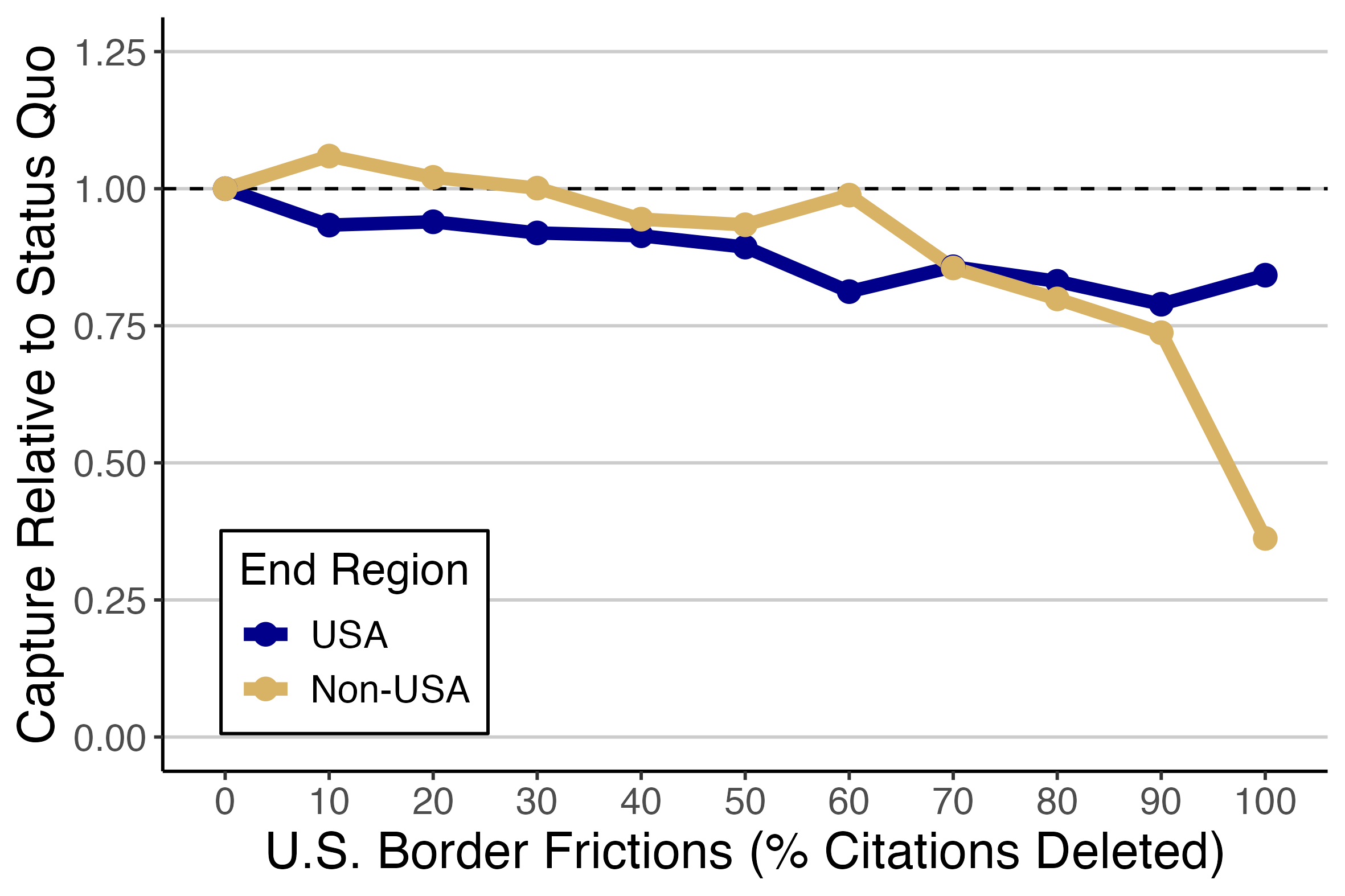}
\end{minipage}\hfill
\begin{minipage}[t]{0.4\textwidth}
\centering
\textbf{B}\\
\includegraphics[width=\textwidth]{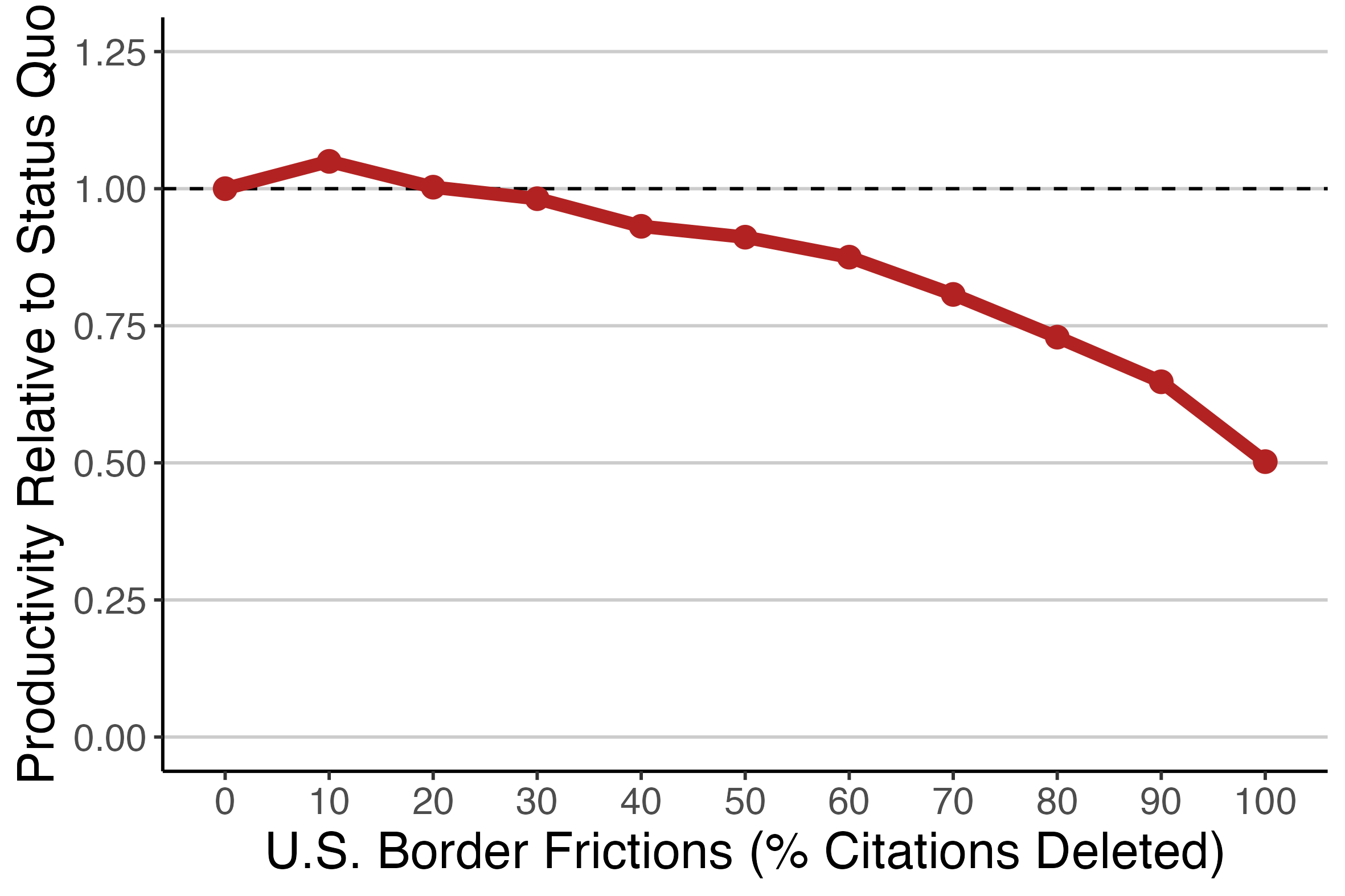}
\end{minipage}

\vspace{1em}

\begin{minipage}[t]{.98\textwidth}
\centering
\textbf{C}\\
\includegraphics[width=\textwidth]{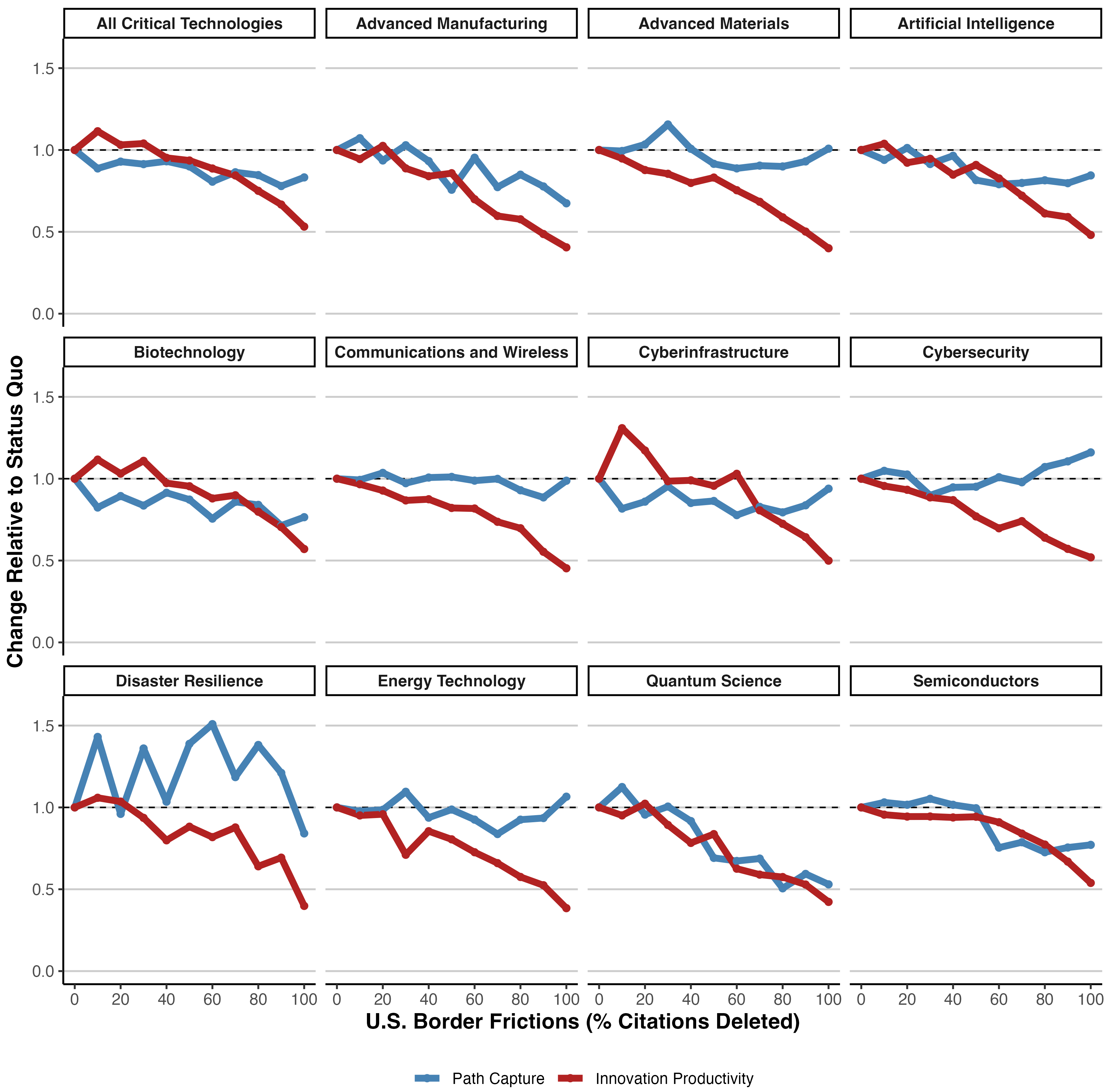}
\end{minipage}

\caption{\linespread{1.0}\selectfont\textbf{Long-Run effects of frictions on moving scientific knowledge across the U.S. border on innovation.} (\textbf{A}) Fewer paths terminate at patents produced both inside and outside the U.S. (\textbf{B}) U.S. innovation productivity declines. (\textbf{C}) U.S. path capture and innovation productivity declines across most NSF-defined Critical Technology Areas.}

\label{fig:fig2}
\end{figure}

\begin{figure}
\centering
\begin{minipage}[t]{0.4\textwidth}
\centering
\textbf{A}\\
\includegraphics[width=\textwidth]{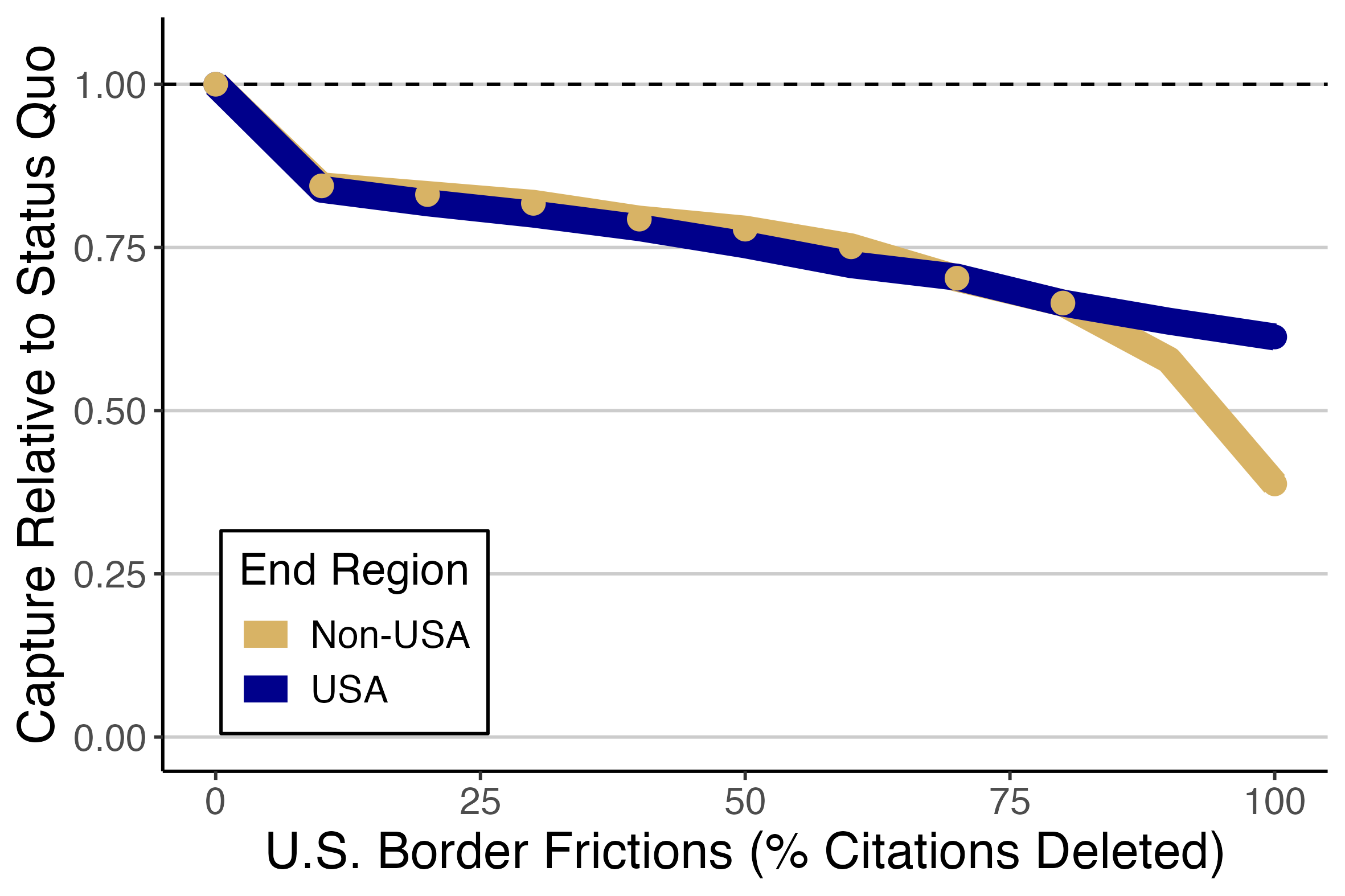}
\end{minipage}\hfill
\begin{minipage}[t]{0.4\textwidth}
\centering
\textbf{B}\\
\includegraphics[width=\textwidth]{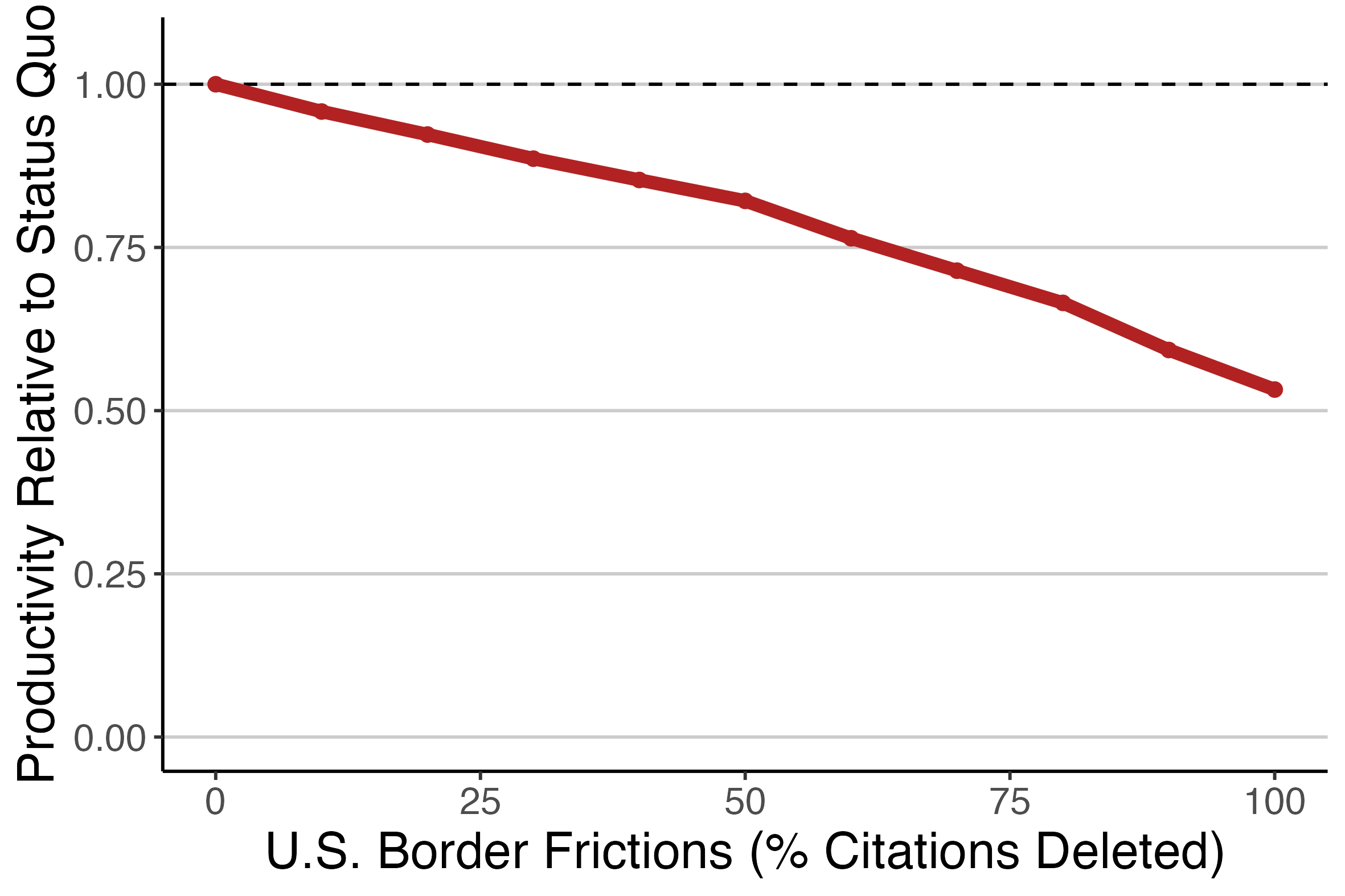}
\end{minipage}

\vspace{1em}

\begin{minipage}[t]{.98\textwidth}
\centering
\textbf{C}\\
\includegraphics[width=\textwidth]{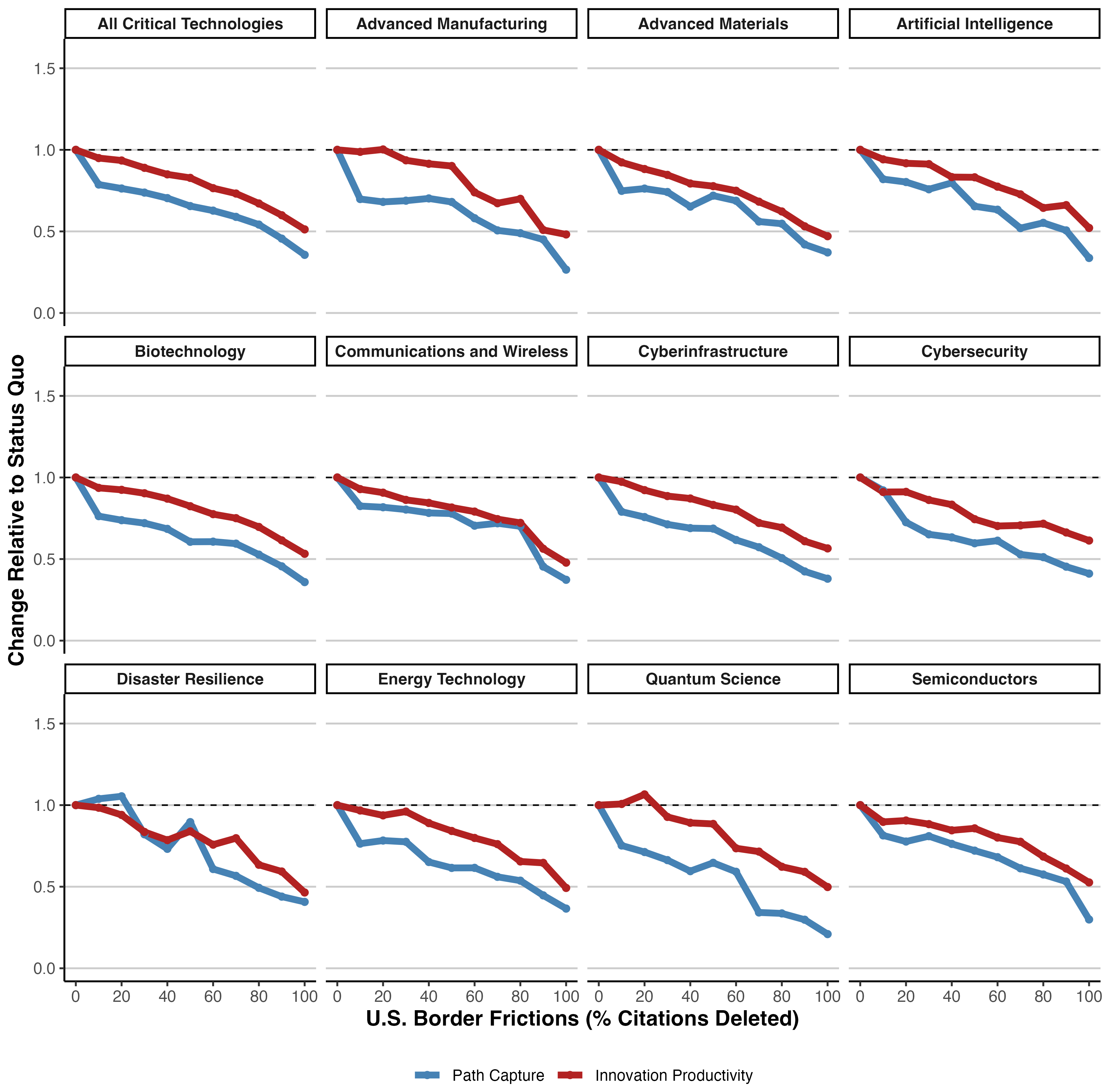}
\end{minipage}

\caption{\linespread{1.0}\selectfont\textbf{Short-run effects of frictions on moving scientific knowledge across the U.S. border on innovation.} (\textbf{A}) Fewer paths terminate at patents produced both inside and outside the U.S. (\textbf{B}) U.S. innovation productivity. (\textbf{C}) U.S. path capture and innovation productivity decrease across all NSF-defined Critical Technology Areas.}
\label{fig:fig_short_run}

\end{figure}


\clearpage
\bibliography{Bib_July12_2025}
\bibliographystyle{sciencemag}

\section*{Acknowledgments}
I thank Olav Sorenson, James Evans, Alan Marco, Megan MacGarvie, and Alexander Furnas for helpful comments, and participants at the Columbia Management, Analytics, and Data conference; the NBER Summer Institute on the Science of Science; Georgia Tech REER; the Entrepreneurship and Innovation Policy Seminar Series; and the Max Planck Institute for Competition and Innovation for valuable feedback.
\paragraph*{Funding:} This research was supported by the Osborne Postdoctoral Fellowship through the Price Center for Entrepreneurship and Innovation at UCLA Anderson School of Management.
\paragraph*{Author contributions:} C.E. designed the research, performed the analysis, and wrote the paper.
\paragraph*{Competing interests:} There are no competing interests to declare.
\paragraph*{Data and materials availability:} All data and code used in this analysis will be deposited in a public repository upon publication. Source data are derived from publicly available datasets including OpenAlex, USPTO PatentsView, the Reliance on Science dataset \cite{marx_reliance_2020}, and SciSciNet \cite{lin_sciscinet_2023}.


\subsection*{Supplementary materials}
Materials and Methods\\
Supplementary Text\\
Figs.~S1 to S17\\
References \textit{(38-\arabic{enumiv})}

\newpage


\renewcommand{\thefigure}{S\arabic{figure}}
\renewcommand{\thetable}{S\arabic{table}}
\renewcommand{\theequation}{S\arabic{equation}}
\renewcommand{\thepage}{S\arabic{page}}
\setcounter{figure}{0}
\setcounter{table}{0}
\setcounter{equation}{0}
\setcounter{page}{1}

\begin{center}
\section*{Supplementary Materials for\\ \scititle}

Christopher~Esposito$^\ast$\\
\small$^\ast$Corresponding author. Email: christopher.esposito@anderson.ucla.edu
\end{center}

\subsubsection*{This PDF file includes:}
Materials and Methods\\
Supplementary Text\\
Figures S1 to S17

\newpage

\subsection*{Materials and Methods}

\subsubsection*{Data Sources}

My analysis combines patent records with data on scientific publications. The patent data come from three sources. First, from the USPTO (via PatentsView), I source patent numbers and inventor home address locations. Second, from the Reliance on Science dataset \cite{marx_reliance_2020}, I source citation records on patents made to scientific articles. Third, from PatStat, I source patent family information. I use the patent family records to identify patents that are applied with the three major patent offices (the USPTO, JPO, and EPO), and I restrict my study to these ``triadic'' patents in order to reduce home-market biases.

For data on scientific publications, I also retrieve records from three sources. First, from OpenAlex I source work ID numbers (hereafter called papers), publication years, and affiliation countries. I source this information for all 239,301,199 papers contained in the November 25, 2024 OpenAlex snapshot. For each paper, I source the affiliation country of the last author. When a paper does not list a last author (such as for single-authored works), I source the country of the paper's first author. I source affiliation locations from OpenAlex's \textit{authorship} object (as opposed to the \textit{institutions} object) because the location information contained in the authorship object does not involve disambiguation across institutions. Thus, authorship-sourced country indicators are available for a larger share of the population of OpenAlex papers than are institutional-sourced country indicators.

OpenAlex is a relatively new data repository, and while its open-access nature and broad coverage present unprecedented opportunities for research, concerns have been raised about data quality. These concerns have primarily focused on two aspects of the dataset: the incorrect disambiguation of authors across papers (a data input not used in this study), and the limited coverage of author affiliations, institutional information, and associated geographies (a data input that this study does rely on). With respect to author affiliations and institutions, OpenAlex has recently made substantial progress in improving data accuracy and coverage; for example, more recent versions of OpenAlex disambiguate author institutions by parsing raw affiliation strings from the authorship sections of papers and matching these against Research Organization Registry (ROR)-linked institution records. Despite these improvements, notable data limitations remain; for instance, the \textit{authorships} object in OpenAlex provides country information for the last authors of only 134,564,231 of the 239,301,199 total papers in its repository---a coverage rate of 56\%.

These concerns pertaining to the completeness of affiliation country data are somewhat less problematic in the context of this study, because the sample analyzed here consists of papers that lie on shortest citation paths linking NSF-funded research to downstream patents. Relative to the full population of papers in OpenAlex, papers on these paths are generally produced by prominent scientists affiliated with well-established institutions. Consequently, OpenAlex has better coverage for these papers: the \textit{authorships} object provides affiliation country information for 95\% of the papers in the sample used in this study. Despite this assurance, it is still possible that missing data could bias key findings. Thus, I show robustness of the core results to various ways in dealing with the remaining papers in the sample that lack affiliation country information. In the main analyses, I drop the 5\% of the works that lack last-author country information.

The second source of publication information is Microsoft Academic Graph (MAG). I source paper-to-paper citation records from the final release version of MAG (December 31, 2021). As with OpenAlex, there have been concerns regarding the quality of MAG data, but these concerns largely center around the disambiguation of scientists across papers and the coverage of affiliation information.

Finally, from SciSciNet I source acknowledgments of NSF funding support, at the paper level \cite{lin_sciscinet_2023}. Articles that acknowledge NSF support serve as the ``start points'' when identifying the citation paths that trace from NSF-funded research to downstream patents. To ensure that my dataset does not include non-substantive papers acknowledging NSF support, I restrict the set of NSF-supported articles to those that receive 10 or more forward citations.

\subsubsection*{Analysis of Coverage of Affiliation Country Data}

An important consideration in any empirical study is the quality of the underlying data. A key data input used in this study is information on the country or location of the last authors of OpenAlex papers.

OpenAlex stores two different types of country information at the paper-author level. The first is \textit{affiliation location}, and the second is \textit{institution location}. In OpenAlex, affiliations are the raw text strings listed on papers. Institutions, in contrast, are disambiguated research entities that have been matched to Research Organization Registry (ROR) entries, which contain a running list of 116,000 disambiguated, consistent institutions.

To produce these data, OpenAlex extracts the raw affiliation text for each author on each paper, and stores that data as the authorship affiliation. It then uses a trained ML model to match these strings to ROR, storing any resulting ROR match as the authorship institution. To collect country code information, OpenAlex first extracts country codes for institutions via the matched RORs. It assigns these country codes to both its institutions and to its affiliations. Finally, for the affiliations that are not matched to institutions, OpenAlex geocodes the raw text strings using a geocoding API. Thus, OpenAlex provides more extensive coverage for country information for affiliations than for institutions. See the OpenAlex GitHub repository for details: \url{https://github.com/ourresearch/openalex-guts/blob/main/files-for-datadumps/standard-format/RELEASE_NOTES.txt}.

To maximize data coverage, I use affiliation-based country code information, accessed through OpenAlex's \textit{authorships} object. Specifically, I source the country code for the last author on each paper. When a paper does not have a last author (such as in a single-authored work), I extract the country code for the first author. Therefore, I retain exactly 1 author and no more than one country code per paper. I hereafter refer to these country codes as the ``countries of the papers''.

In its entirety, OpenAlex contains 239,301,199 papers as of its November 25, 2024 snapshot. Using the above-mentioned methods, I extract country information for 134,564,231 (or 56\%) of the full population.

While country information is only available for 56\% of the full population, coverage is considerably higher for the sample analyzed in this study. This is because the analyzed sample consists of papers that fall on the shortest citation paths connecting NSF funded research to downstream patents. In particular, OpenAlex provides country information for 32,295,115 of the 45,857,172 papers that can be linked to downstream patents (70.4\% coverage), and it contains country information for 650,644 of the 683,222 papers that fall on the shortest paths from NSF-funded research to the nearest patent (95.2\% coverage). Thus, the coverage rate of papers analyzed in the sample is relatively high.

In the main analysis, I omit papers lacking country information from the sample. I omit these papers \textit{before} calculating the shortest paths between NSF-funded articles and patents. Thus, if the shortest path between an NSF-funded paper and patent contains a paper lacking country information, I allow the path to reconfigure along its shortest remaining paths.

\begin{figure}
\centering
\includegraphics[width=0.95\textwidth]{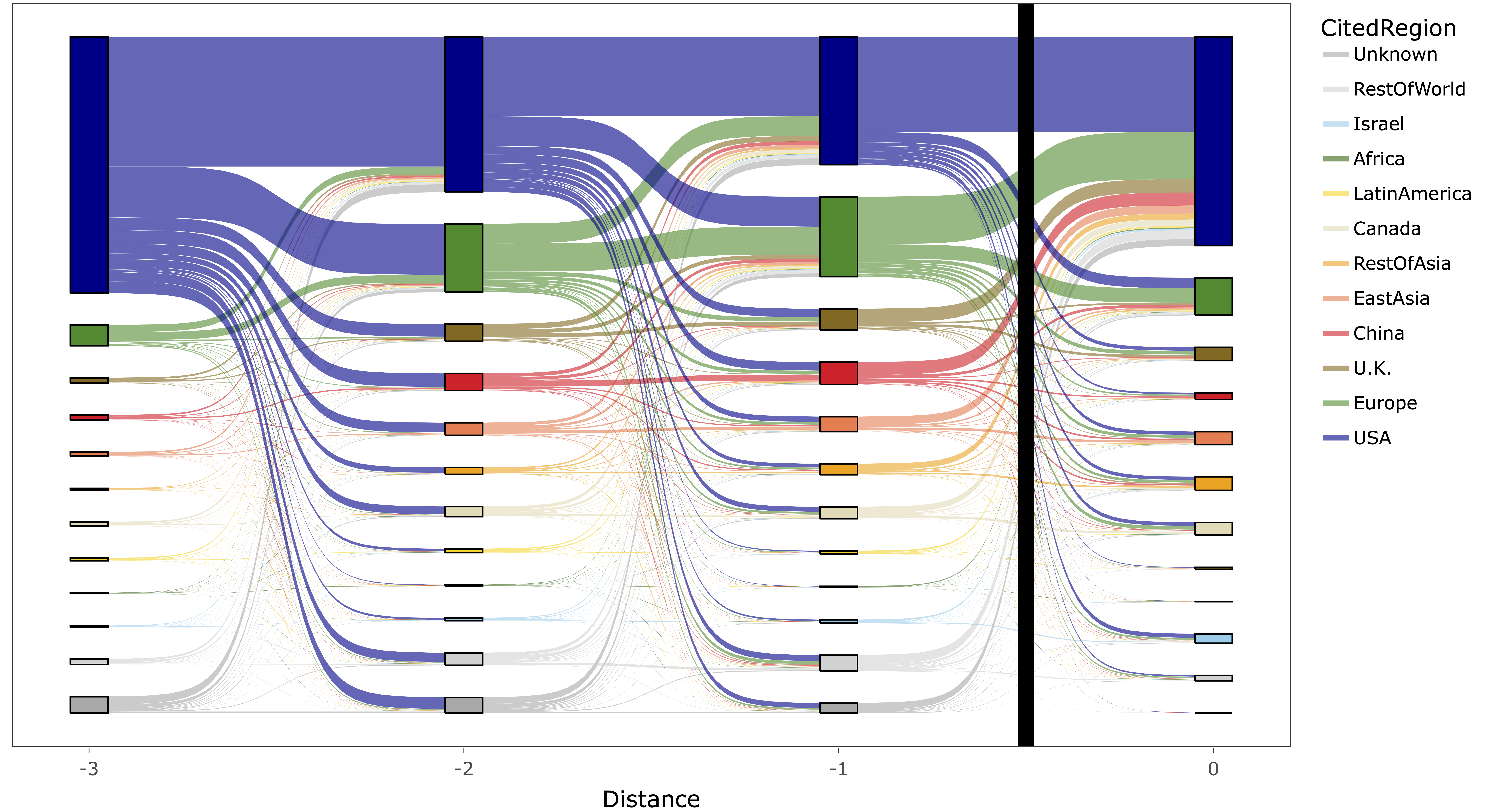}
\caption{\textbf{Flow Diagram of Paths Including Papers with Unknown Country Affiliations.}}
\label{fig:s_flow_no_unknown}
\end{figure}

In Figure~\ref{fig:s_flow_no_unknown}, I present a flow map illustrating the regional trajectories of paths connecting NSF-funded research to downstream patents, retaining papers with unknown affiliation countries in the dataset. This figure provides a transparent view of the role such ``unknown'' country papers play in the overall pattern. As shown, relatively few papers along these paths lack country information, and their presence is inconsequential to the broader dynamic: paths typically originate in the U.S., shift to other regions as they approach the patent boundary, and return to the U.S. at the point of patenting. I therefore conclude that papers with missing country data do not meaningfully affect the overall conclusions of the analysis.

\subsubsection*{Methods for Constructing Shortest Citation Paths}

I construct a shortest-path network that links NSF-supported papers to their nearest patents. To do so, I begin with USPTO records of triadic patents, front-page citations from patents to scientific articles \cite{marx_reliance_2020}, and citations from scientific articles to other scientific articles (OpenAlex). I call this complete network of all front-page patent-to-paper and all paper-to-paper citations the ``full'' citation network.

Next, I identify the scientific papers directly cited by the triadic patents. These direct citations are reflected by the patents (black squares) in Figure~\ref{fig:fig1}A of the main text. In the figure, the patents receive inward-pointing edges from papers, Papers 1, 2, and 3. Because these papers are directly cited by patents, they are positioned at a value of $-1$ on the $x$ axis, indicating that the shortest path length from these papers to the nearest patent is 1.

After identifying the papers that are directly cited by patents, I identify the papers that are indirectly cited by patents. To do so, I begin with the papers that are directly cited and record the papers that they cite: Papers 4, 5, and 6, as shown in Figure~\ref{fig:fig1}A. These papers are located at a value of $-2$ on the $x$ axis, because the shortest length path from them to the nearest patent is 2.

Next, I identify the papers at shortest length distance 3 from the nearest patent. To do so, I extract the citations made by papers at $Distance = -2$. In Figure 1, there are two papers cited by papers at $Distance = -2$: Papers 5 and 7. However, Paper 5 was also cited by Paper 2. Since Paper 5 is cited by both Paper 4 and Paper 2, and Paper 2 lies on a shorter path to the patent, the citation from Paper 5 to Paper 4 does not fall on the shortest path. Therefore, I omit the citation running from Paper 5 to Paper 4 when computing the shortest paths.

I repeat the above steps until I arrive at scientific papers that are at $Distance=-10$. There are a total of 31,278,142 scientific papers in the citation graph, linked to 436,604 patents.

After constructing the shortest-path network, I extract the paths of the network that begin with papers that acknowledge NSF support. There are 792,014 total articles that acknowledge NSF support in the SciSciNet dataset, of which 499,763 receive 10 or more forward citations. I conduct walks along the shortest-path network from each of these 499,763 articles, moving forward across the citation network until I reach a patent. When more than one equally-short walk is possible between an NSF-supported article and a patent, I randomly select one of them. In the status quo scenario, in which there are no new barriers to knowledge flow across the U.S. border, 360,419 of the 499,763 NSF-funded articles (72\%) are linked to 50,813 downstream patents.

\subsubsection*{Methods for Simulating Knowledge Flow Barriers at the U.S. Border}

To simulate barriers to knowledge flows crossing the U.S. border, I first identify citations in the full citation network where the last author of the cited paper is outside the U.S. and the last author of the citing paper is in the U.S., or the last author of the cited paper is in the U.S. and the last author of the citing paper is outside the U.S. Next, I remove each of these cross-border citations with a set probability, producing multiple restricted networks. The 0\% exclusion network removes no citations that cross the U.S. border; I call this network the status quo. The restricted networks remove citations crossing the U.S. border with 10\%, 20\%, \ldots, up to 100\% probability. After constructing these networks, I identify the shortest paths connecting NSF-supported articles to patents in each of them using the steps shown in the main text (also see Fig.~\ref{fig:fig1}A).

\subsubsection*{Methods for Down-Sampling Outstanding Paths}

To forecast where the outstanding paths (NSF-funded articles that have not yet been linked to a patent) will be captured, I first down-sample the full set of these outstanding start-points. I conduct this down-sampling because not all of the outstanding paths will eventually become linked to a patent. To down-sample the outstanding paths, I use a hazards model to probabilistically remove outstanding paths based on their age and the distribution of the age of realized paths as of the year that they first become linked to patents.

\begin{figure}
\centering
\includegraphics[width=0.7\textwidth]{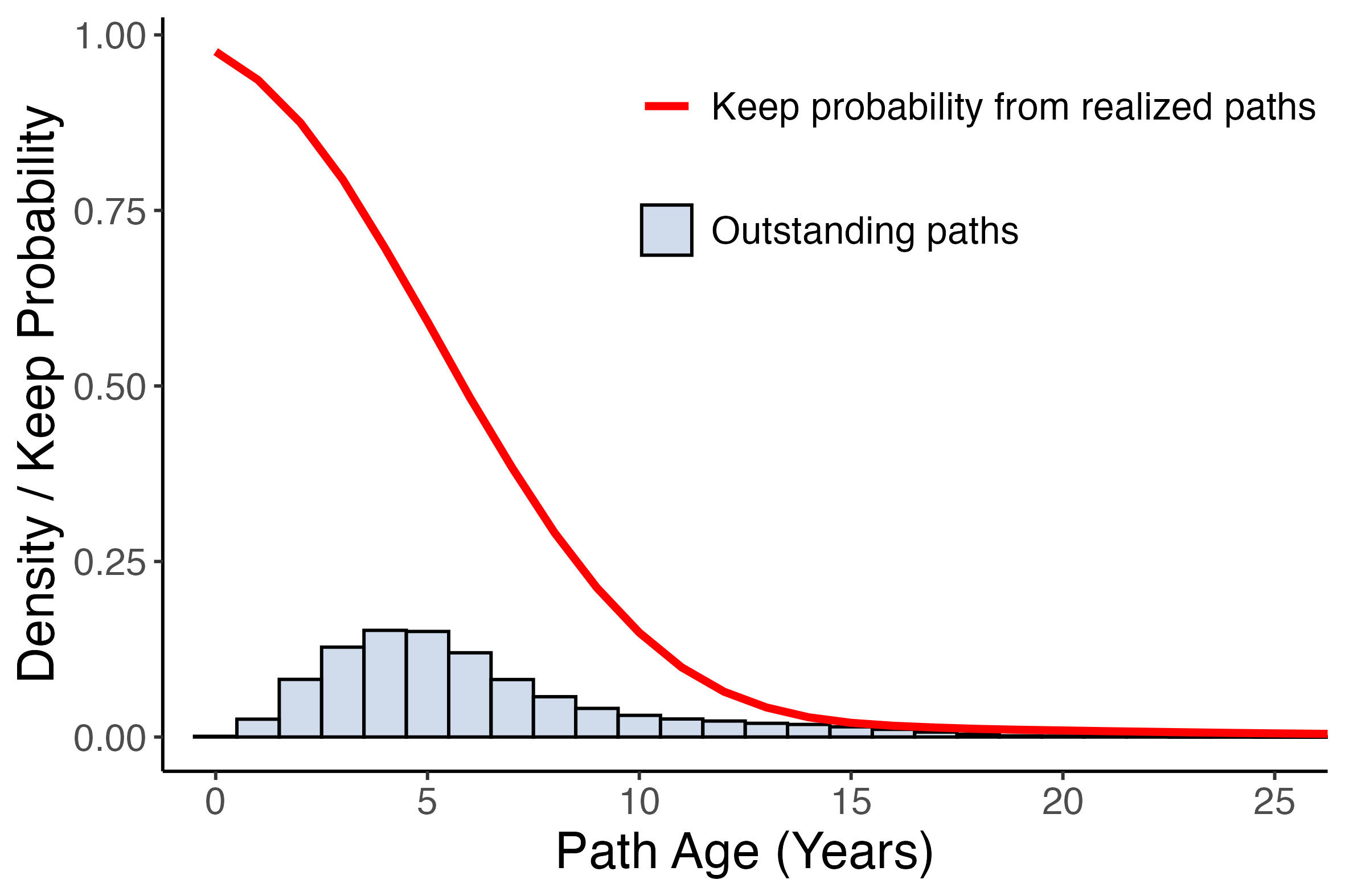}
\caption{\textbf{Down-sampling outstanding paths based on the keep probability.} The keep probability is calculated using the inverse hazards function of realization with a patent based on its age in years, calculated using the realized paths. The resulting keep probability is used to down-sample the outstanding paths.}
\label{fig:s_downsampling}
\end{figure}

More specifically, I first identify all realized paths (those that have already linked to a patent), and record the number of years between the publication of their start-paper and their downstream patent (i.e.\ years-to-realization). Next, I identify all outstanding paths and calculate their age as of 2021 (the last year covered by my patent data). Finally, I down-sample outstanding paths based on their age, using the empirical distribution of years-to-realization from the realized paths. For example, 99\% of realized paths were linked to patents by age 17, so outstanding paths older than 17 years are unlikely to connect to a patent in the future. These outstanding paths are probabilistically dropped from the sample based on the observed hazards function from the realized paths. Figure~\ref{fig:s_downsampling} shows how the outstanding paths are down-sampled based on the keep probability calculated from the realized paths.

The down-sampling process reduces the number of outstanding paths from 194,452 to 100,899. The median path age of the surviving outstanding paths (bench-marked to 2021, the final year of the patent-to-paper citation dataset) is 6 years.

\subsubsection*{Methods for Assigning Outstanding Paths to (StartingD, CurrentD, CurrentRegion) Cells}

After down-sampling the outstanding paths, I assign three features to each: the start paper's distance to the nearest patent (\textit{StartingD}), the current distance of the path in 2021 (\textit{CurrentD}), and the current region where the path is being advanced (\textit{CurrentRegion}). I make these assignments using a single-stage probabilistic draw that conditions jointly on the path's observed age and its projected total age-at-realization.

Let $a$ be the observed path age (as of 2021) and $\hat{A}$ the projected age-at-realization. Let $S \in \mathcal{S}$, $C \in \mathcal{C}$, and $R \in \mathcal{R}$ denote the possible values of \textit{StartingD}, \textit{CurrentD}, and \textit{CurrentRegion}, respectively. From the realized paths, I estimate the empirical joint distribution
\[
p_{SCR|a,\hat{A}} = \Pr(S = s, C = c, R = r \mid \text{age} = a, \ \text{age-at-realization} = \hat{A}),
\]
where $p_{SCR|a,\hat{A}}$ is the observed frequency of $(s,c,r)$ among realized paths with the same $(a,\hat{A})$.

I then assign each outstanding path to $(S, C, R)$ cells by drawing
\[
(S, C, R) \sim p_{SCR|a,\hat{A}},
\]
so the probability of assigning an outstanding path to a cell $(s,c,r)$ matches its empirical share in the realized-path data for the corresponding $(a,\hat{A})$ bin.

To project the age-at-realization for outstanding paths, I sample residual durations from the realized paths' distribution of (age-at-realization $-$ current age), conditional on path age ($a$). Adding the residual years to $a$ yields $\hat{A}$, ensuring that the simulated distribution of total ages-at-realization matches that of the realized paths.

This single-stage approach assigns all three features (StartingD, CurrentD, CurrentRegion) in one probabilistic draw, preserving relationships observed in the realized paths. For example, longer-to-realize paths tend to start farther from the patent boundary and, conditional on starting point, older paths tend to be closer to the boundary.

\begin{figure}
\centering
\includegraphics[width=0.7\textwidth]{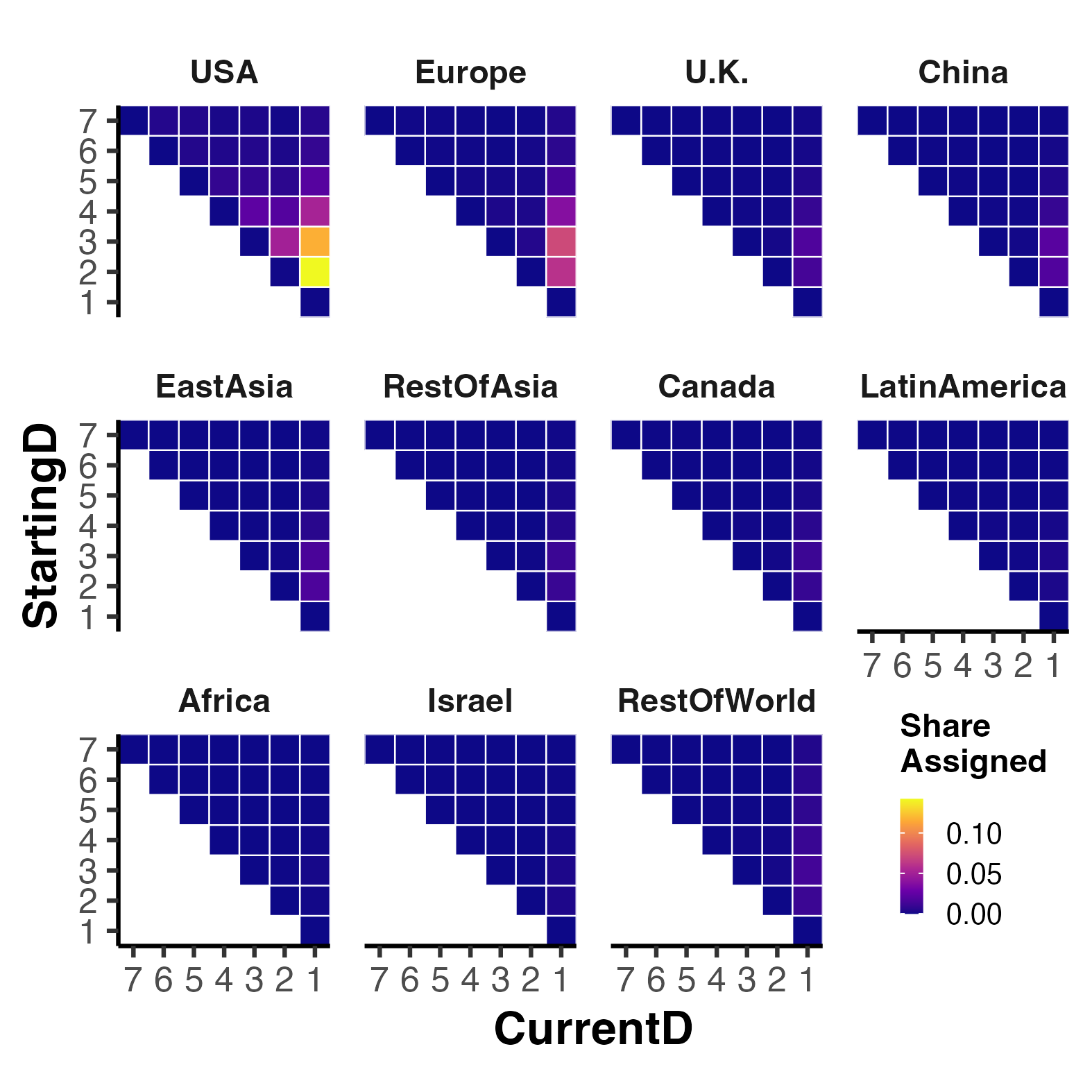}
\caption{\textbf{Assignment of Outstanding Paths to (StartingD, CurrentD, CurrentRegion) Cells.}}
\label{fig:s_outstanding_path_assignment}
\end{figure}

Figure~\ref{fig:s_outstanding_path_assignment} shows the share of outstanding paths assigned to each cell. Most outstanding paths are assigned to cells in the USA, generally with CurrentD values of $-1$ or $=2$, and StartingD values of $-2$, $-3$, or $-4$. A smaller but significant share are assigned to cells in Europe, the U.K., and China, as indicated by the shading in Figure~\ref{fig:s_outstanding_path_assignment}.

\begin{figure}
\centering
\includegraphics[width=0.5\textwidth]{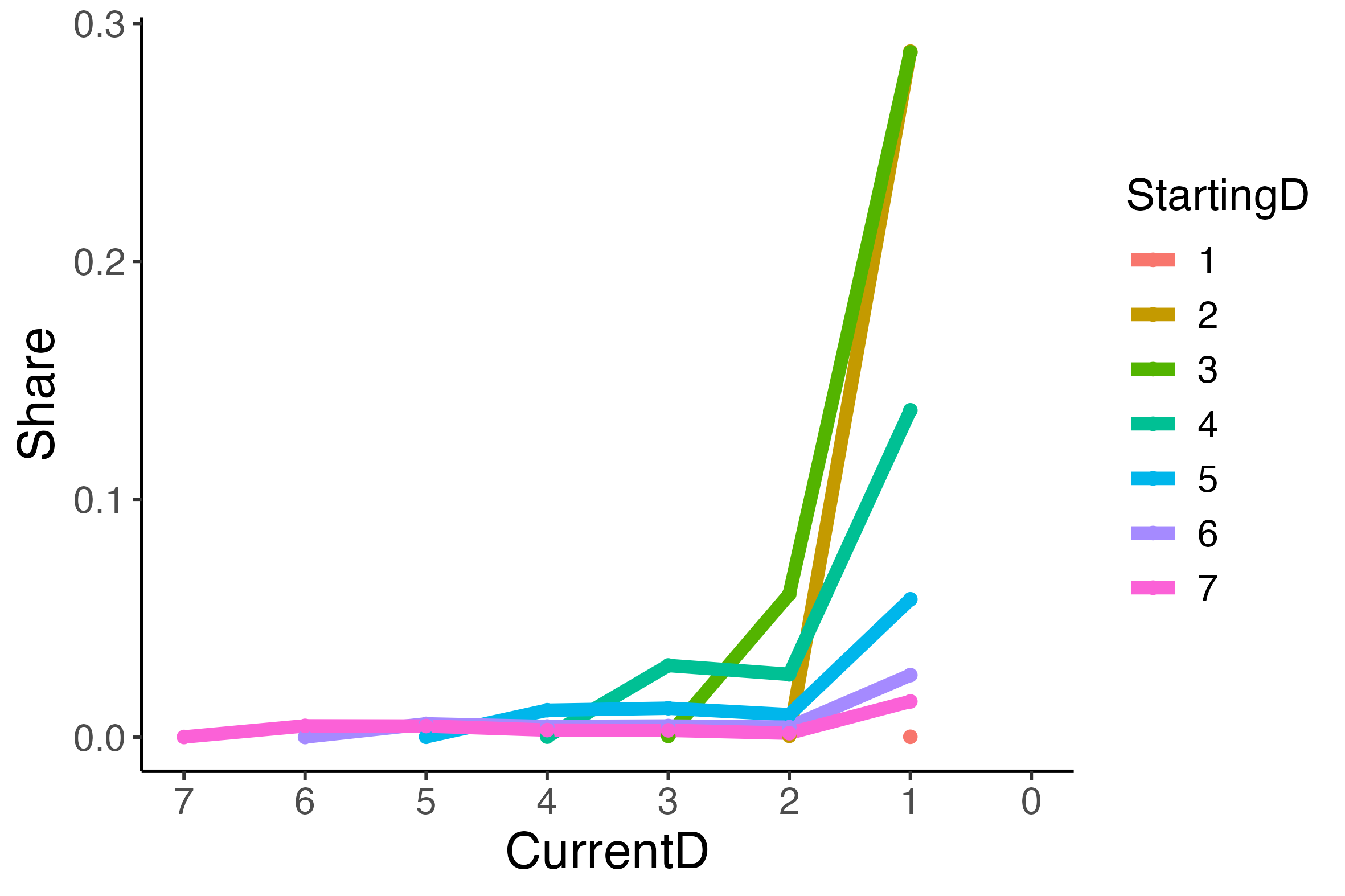}
\caption{\textbf{Assignment of outstanding paths to StartingD and CurrentD values.}}
\label{fig:s_outstanding_path_assignment_d}
\end{figure}

To show the assignments at higher levels of aggregation, Figure~\ref{fig:s_outstanding_path_assignment_d} pools across regions to show the total share of outstanding paths assigned to specific StartingD and CurrentD values: 0.01\% of paths are assigned to $StartingD = -1$, 14.5\% to $StartingD = -2$, 30.6\% to $StartingD = -3$, 24.5\% to $StartingD = -4$, and smaller shares to higher StartingD values. For CurrentD, 50.2\% are assigned to $CurrentD = -1$, 27.6\% to $CurrentD = -2$, with the rest at higher values.

\begin{figure}
\centering
\includegraphics[width=0.5\textwidth]{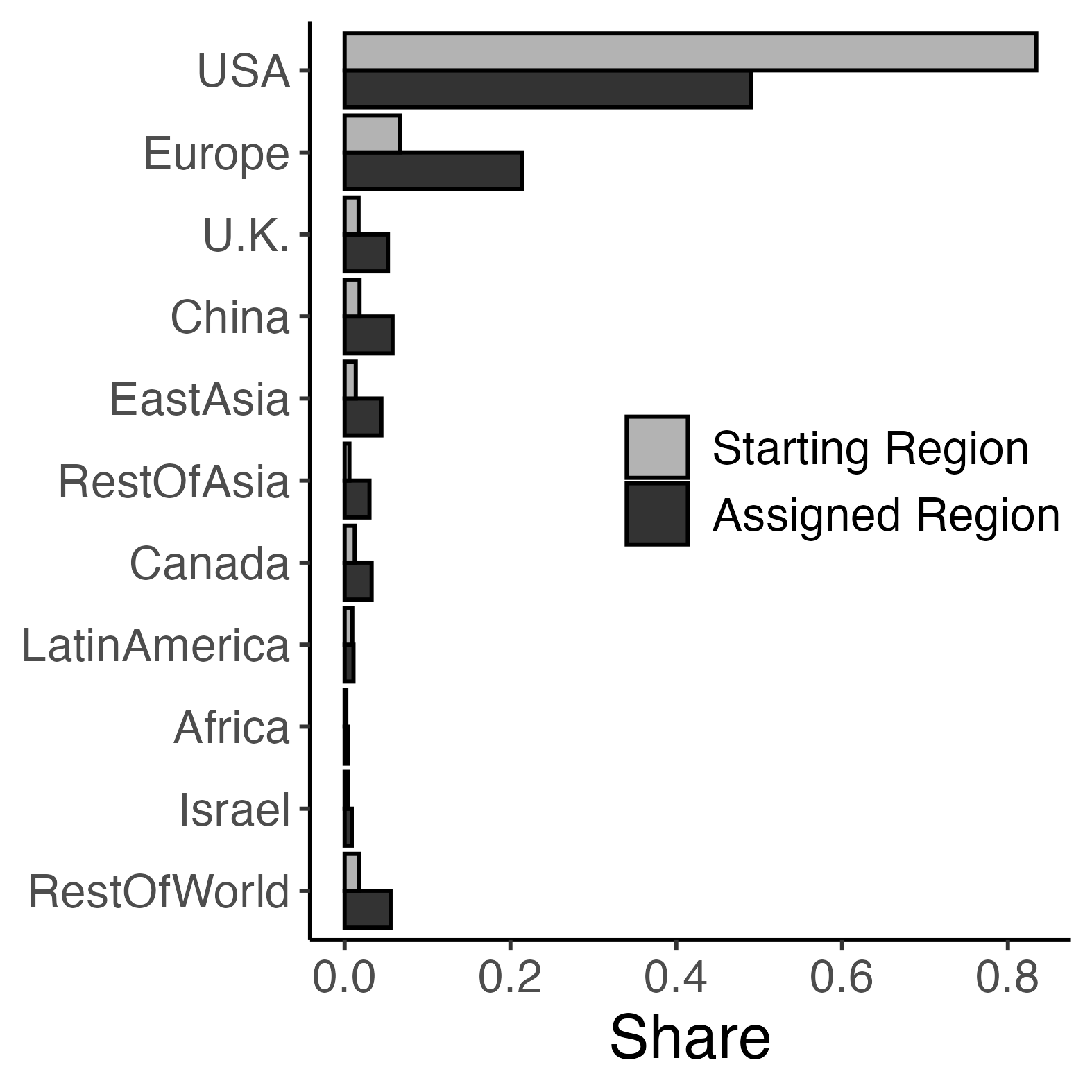}
\caption{\textbf{Assignment of outstanding paths to regions.}}
\label{fig:s_outstanding_path_assignement_regions}
\end{figure}

Finally, Figure~\ref{fig:s_outstanding_path_assignement_regions} shows the resulting regional assignments. Regions with many realized-path papers in a given cell are proportionally more likely to receive outstanding paths assigned to that cell, which often differs from the region of the starting paper, especially when $CurrentD < StartingD$.

\subsubsection*{Methods for Computing Transition Probabilities}

After assigning outstanding paths to $(\mathrm{StartingD}, \mathrm{CurrentD}, \mathrm{CurrentRegion})$ cells, I project their likely outcomes. In particular, I predict whether and where they ultimately terminate in a patent along their shortest path, and the distance of that termination from the starting point. I estimate these outcomes under two conditions: (i) in the absence of barriers to international knowledge flows (status quo), and (ii) in the presence of such barriers.

For the status quo case, I use the realized paths to compute, for each $(\mathrm{StartingD} = s, \mathrm{CurrentD} = d, \mathrm{CurrentRegion} = r)$ cell, the empirical probability that a path terminates in a patent located in region $r \in \mathcal{R}$ at distance $d \in \mathcal{D}$.

For the restricted-network cases (10\%, 20\%, \ldots, 100\% reductions in cross-border knowledge flows with the U.S.), I again use the realized paths to compute, for each $(s, d, r)$ cell, the empirical probabilities that a path terminates in a patent in the same region $r$ as before with the same $\mathrm{CurrentD} = d$; terminates in a patent in a different region $r' \neq r$ or at a greater distance $d' > d$; or stalls (does not terminate in a patent). These three outcomes are mutually exclusive and exhaustive.

\subsubsection*{Methods for Simulating Outcomes of Outstanding Paths}

After computing the transition probabilities from realized paths, I simulate the outcomes of outstanding paths by drawing randomly from the transition probabilities corresponding to their assigned $(s, d, r)$ cell. This simulation is performed using probabilities from both the unrestricted and restricted networks, producing comparable projected outcomes across the full range of possible U.S. cross-border flow restrictions.

\subsubsection*{Assignment of Countries to Regions}

Countries are aggregated to regions based on their location, embeddedness in supranational bodies, and geopolitical significance. The regional definitions are as follows:

\textbf{Europe} = the 27 member states of the European Union, Switzerland, and Norway. \textbf{East Asia} = South Korea, Japan, Taiwan. \textbf{Rest of Asia} = Afghanistan, Armenia, Azerbaijan, Bahrain, Bangladesh, Bhutan, Brunei, Cambodia, Georgia, Indonesia, India, Iran, Iraq, Jordan, Kazakhstan, Kuwait, Kyrgyzstan, Laos, Lebanon, Malaysia, Maldives, Mongolia, Myanmar, Nepal, North Korea, Oman, Pakistan, Palestine, Philippines, Qatar, Saudi Arabia, Singapore, Sri Lanka, Syria, Tajikistan, Thailand, Timor-Leste, Turkmenistan, United Arab Emirates, Uzbekistan, Vietnam, Yemen. \textbf{Latin America} = Argentina, Belize, Bolivia, Brazil, Chile, Colombia, Costa Rica, Cuba, Dominica, Dominican Republic, Ecuador, El Salvador, Grenada, Guatemala, Guyana, Haiti, Honduras, Jamaica, Mexico, Nicaragua, Panama, Paraguay, Peru, Saint Kitts and Nevis, Saint Lucia, Saint Vincent and the Grenadines, Suriname, Trinidad and Tobago, Uruguay, Venezuela, Bahamas, Barbados. \textbf{China} = Mainland China, Hong Kong, Macau. \textbf{Africa} = Algeria, Angola, Benin, Botswana, Burkina Faso, Burundi, Cameroon, Cape Verde, Central African Republic, Chad, Comoros, Republic of the Congo, C\^ote d'Ivoire, Democratic Republic of the Congo, Djibouti, Egypt, Equatorial Guinea, Eritrea, Eswatini, Ethiopia, Gabon, The Gambia, Ghana, Guinea, Guinea-Bissau, Kenya, Lesotho, Liberia, Libya, Madagascar, Malawi, Mali, Mauritania, Mauritius, Morocco, Mozambique, Namibia, Niger, Nigeria, Rwanda, S\~ao Tom\'e and Pr\'incipe, Senegal, Seychelles, Sierra Leone, Somalia, South Africa, South Sudan, Sudan, Tanzania, Togo, Tunisia, Uganda, Zambia, Zimbabwe. The \textbf{United States}, \textbf{Canada}, \textbf{U.K.}, and \textbf{Israel} are defined as independent regions. The \textbf{Rest of World} region includes all countries not mentioned in the above lists.

\subsubsection*{Assignment of Patents to NSF Critical Technology Areas}

To assign patents to NSF-defined Critical Technology Areas (CTAs), I constructed a crosswalk between Cooperative Patent Classification (CPC) codes and CTAs using a combination of structured classification data and large language model (LLM) assistance.

I began by extracting CPC main groups from the CPC classification system and constructing full-text descriptions for each group by concatenating class-, subclass-, and group-level titles. These descriptions provided a standardized representation of the technological content of each CPC category.

I then mapped CPC groups to CTAs using an LLM. The model was prompted to assign each CPC group to the most relevant CTA, given both the CPC group description and an NSF-provided dictionary of keywords associated with each CTA (see \cite{wu_chinas_2024}). Including these keywords anchored assignments in NSF's official taxonomy and improved consistency across related technological domains. I subsequently reinserted the initial CPC--CTA crosswalk into the model for a second pass to identify and correct low-confidence or potentially ambiguous assignments.

Following this automated classification, I manually reviewed and corrected a small number of misclassifications, with particular attention to over-assignment into quantum technologies.

I then merged the CPC--CTA crosswalk to patent-level CPC data. For each patent, I retained only the CTA associated with its primary CPC classification (i.e., the first-listed CPC code), ensuring that each patent was assigned to at most one CTA. This produced a one-to-one mapping between patents and CTAs that could be consistently linked to citation paths in the analysis.

This approach combined the hierarchical structure of CPC classifications with LLM-based semantic matching, enabling scalable and interpretable assignment of patents to policy-relevant technology categories.

\newpage

\subsection*{Supplementary Text}

\subsubsection*{Motivating Discussion}

New inventions often build on diverse sets of scientific ideas, many of which originate outside the focal domain or geography of application. Because invention entails the creation of novelty, it benefits from access to a wide and varied knowledge base, which expands the set of possible combinations and increases the likelihood of discovering useful new assemblies \cite{weitzman1998,fleming_recombinant_2001}. To find new knowledge inputs, scientists and inventors often benefit from search across geographical borders, where different sets of preferences and demand nurture distinctive areas of expertise. For example, in the case of CRISPR, foundational advances were not only made by medical researchers attempting to improve human health outcomes, but also by scientists in Dang\'e-Saint-Romain, a dairy-processing region in France, where work was underway to address microbial contamination problems in yogurt \cite{barrangou_2007}. The fundamental insight developed in the lab, that inserting new sequences into bacterial CRISPR arrays could generate targeted phage resistance in yogurt cultures, was later built on by U.S.-based inventors to re-engineer the system into a functioning gene-editing technology \cite{zhang_crispr-cas_2014,doudna_methods_2019}. This case illustrates how the extent and differentiation of international markets fosters a globally-heterogeneous stock of scientific knowledge and enables innovation to advance in unforeseen ways.

Yet while scientific knowledge is often treated as a global and public good, its value for innovation depends on the ability to access, absorb, and apply it. Firms and countries alike require complementary assets, organizational capabilities, and institutional infrastructures to translate scientific inputs into commercial outcomes \cite{teece_1986,furman_2002,azoulay_2025}. In the U.S., these complementary assets are highly concentrated in a small number of regional innovation hubs, including the San Francisco Bay Area and Greater Boston, where dense networks of universities, venture capital, entrepreneurial firms, and tech transfer institutions facilitate the conversion of scientific research into marketable technologies \cite{bikard_bridging_2020,samila_venture_2010,samila_venture_2011}.

From a national strategy perspective, the United States' ability to access and commercialize global science confers major advantages. By funding and publishing foundational research at scale, the U.S. disproportionately shapes global research priorities in ways aligned with its strategic interests \cite{wu_chinas_2024}. In addition, by concentrating many of the world's complementary assets for commercialization, the U.S. captures a disproportionate share of the downstream benefits of scientific discovery, including monopoly rents, tax revenues \cite{storper_2015}, and technologies that advance national security \cite{miller_chip_2022}. Finally, because foreign researchers produce research that is ultimately commercialized in the U.S., foreign governments, firms, and funding agencies effectively subsidize U.S. innovation.

However, this arrangement also reduces U.S. economic autonomy. By relying heavily on global knowledge flows, U.S. innovation becomes vulnerable to disruptions in international scientific exchange. Such disruptions often result from restrictions in the movement of people across geographical boundaries. Because many ideas have a tacit dimension, their transmission is aided by face-to-face communication \cite{polanyi_tacit_1966,storper_2004}. Quasi-experimental evidence shows that geographical proximity affects the diffusion of scientific and technological ideas. For example, the emigration of Jewish inventors from Nazi Germany to the U.S. accelerated innovation in the treated domains \cite{moser_2014}, and the fall of the Soviet Bloc diffused foreign mathematical ideas to the U.S. \cite{teodoridis_creativity_2019}. The inverse relationship is true as well: evidence from premature deaths of inventors shows that the unexpected removal of individuals from geographical regions reduces local knowledge diffusion \cite{balsmeier_isolating_2023}. Even relatively mundane barriers to in-person knowledge exchange negatively impact diffusion. For instance, scheduling conflicts at computer science conferences reduce the likelihood of citing a presented paper by 62.5\% \cite{teplitskiy_2024}. If the U.S. borders do not remain open to the movement of scientists and inventors, they may also close off the movement of ideas, and affect the ability for the U.S. economy to commercialize global science.

Measures have been enacted that impede the movement of people across the U.S. border. Some of these measures are explicit, such as changes in visa regimes and limits on international collaboration for federally-funded research \cite{chatterji_how_2025,dhs_2025}. Other measures are not explicit, but still affect the propensity for scientists and inventors to travel across the border. For example, anti-foreigner rhetoric decreases the sense of security of foreign scientists working in the U.S., leading to out-migration and diminished participation at international research conferences \cite{wu_chinas_2024,naddaf_scientific_2025}. Interventions affecting the diffusion of knowledge to the U.S. have also been imposed by foreign governments. For instance, China's Thousand Talents Program re-domiciled leading Chinese scientists \cite{shi_has_2023}, while its China Journal Excellence Action Plan, launched in 2019, aims to elevate 400 domestically-published scientific journals, fostering a more self-contained research ecosystem \cite{owens_2024}. Both policies may reduce U.S. access to Chinese science. In addition, in the changing geopolitical environment, traditional U.S. allies come to see the U.S.' reliance on global science as an opportunity for leverage. For example, in early 2025, Canada responded to new U.S. tariffs by enacting targeted import duties on Republican-voting districts \cite{djuric_2024}. If foreign governments come to recognize the extent of U.S. dependence on their scientific outputs and develop mechanisms to control their diffusion, then export restrictions on scientific knowledge could develop into negotiating devices for international diplomacy---analogous to tariffs on physical goods, but with a potentially different balance of power.

Existing research has documented the international character of scientific collaboration using co-authorship networks, researcher mobility, and citation patterns across borders \cite{wagner_2005}. However, these approaches often emphasize the presence of collaboration, rather than the power or influence that one country may have over another \cite{chatterji_how_2025}. While recent studies have documented the growing use of Chinese science in U.S. patenting \cite{wang_contribution_2024} and shown the growing leadership position of Chinese scientists involving collaborators in the U.S. \cite{wu_chinas_2024}, prior research has not measured the dependence of U.S. innovation on global science, nor simulated the resilience of U.S. innovation to geopolitical disruption.

This paper addresses that gap by tracing the citation pathways from NSF-funded science to downstream patents and simulating scenarios in which cross-border knowledge flows are disrupted. These simulations offer a quantitative stress-test of the U.S. scientific supply chain, revealing how innovation outcomes change when access to foreign science is reduced. In doing so, the paper contributes to an emerging view of science not only as a source of knowledge, but as a strategic resource that is globally distributed, deeply interdependent, and vulnerable to policy shocks \cite{chatterji_how_2025}. By connecting theories and methods from innovation strategy, economic geography, and the economics of science, it offers a new empirical framework for evaluating the national consequences of scientific de-globalization.

\subsubsection*{Impacts of Border Restrictions on Capture and Productivity, Raw Values}

\begin{figure}
\centering
\begin{minipage}[t]{0.48\textwidth}
\centering
\textbf{A}\\
\includegraphics[width=\textwidth]{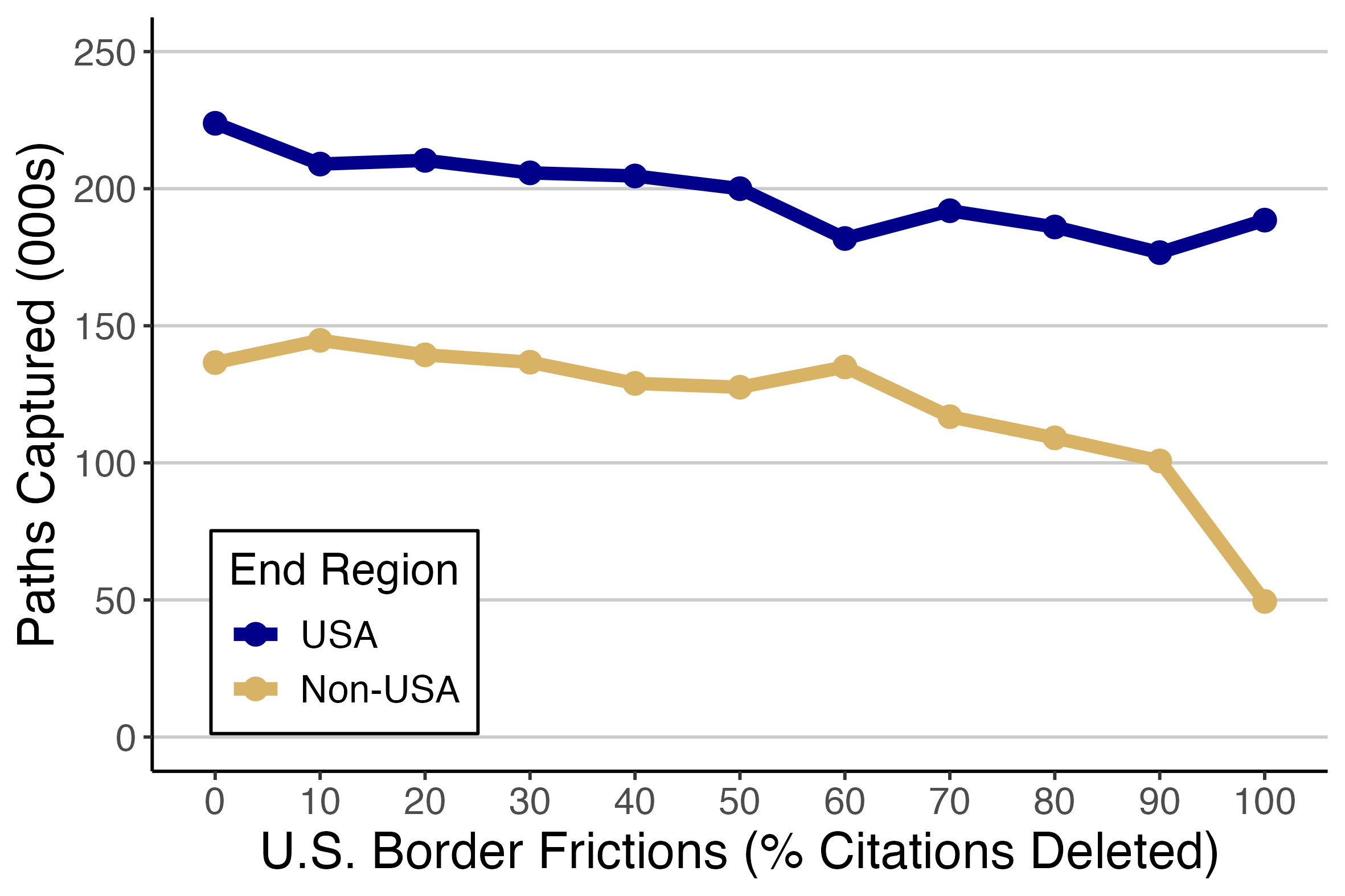}
\end{minipage}\hfill
\begin{minipage}[t]{0.48\textwidth}
\centering
\textbf{B}\\
\includegraphics[width=\textwidth]{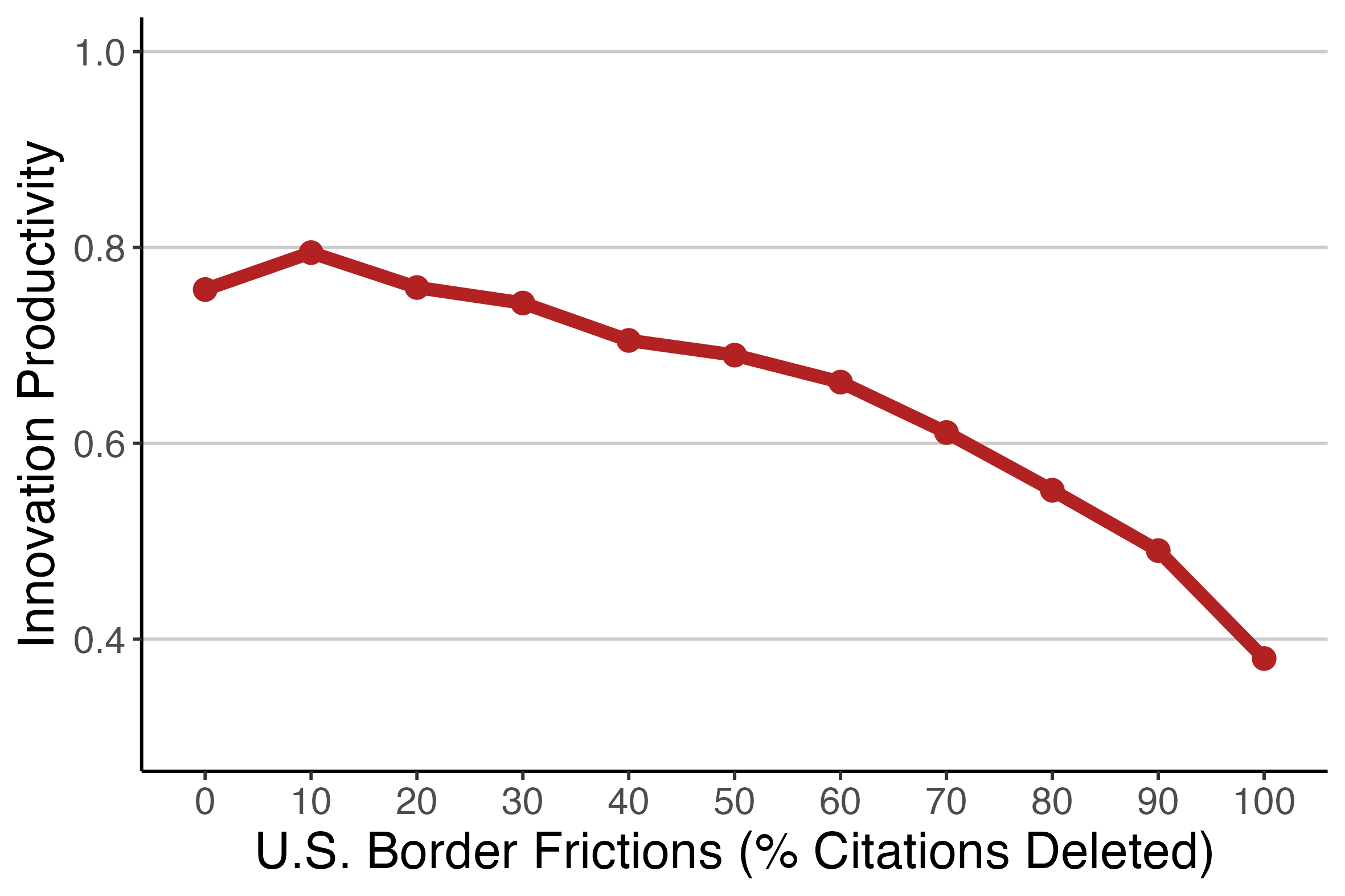}
\end{minipage}
\caption{\textbf{Long-Run Effects of Knowledge Flow Restrictions, Raw Values.} (\textbf{A}) Paths captured. (\textbf{B}) U.S. innovation productivity.}
\label{fig:s_capture_absolute_longrun}
\end{figure}

\begin{figure}
\centering
\begin{minipage}[t]{0.48\textwidth}
\centering
\textbf{A}\\
\includegraphics[width=\textwidth]{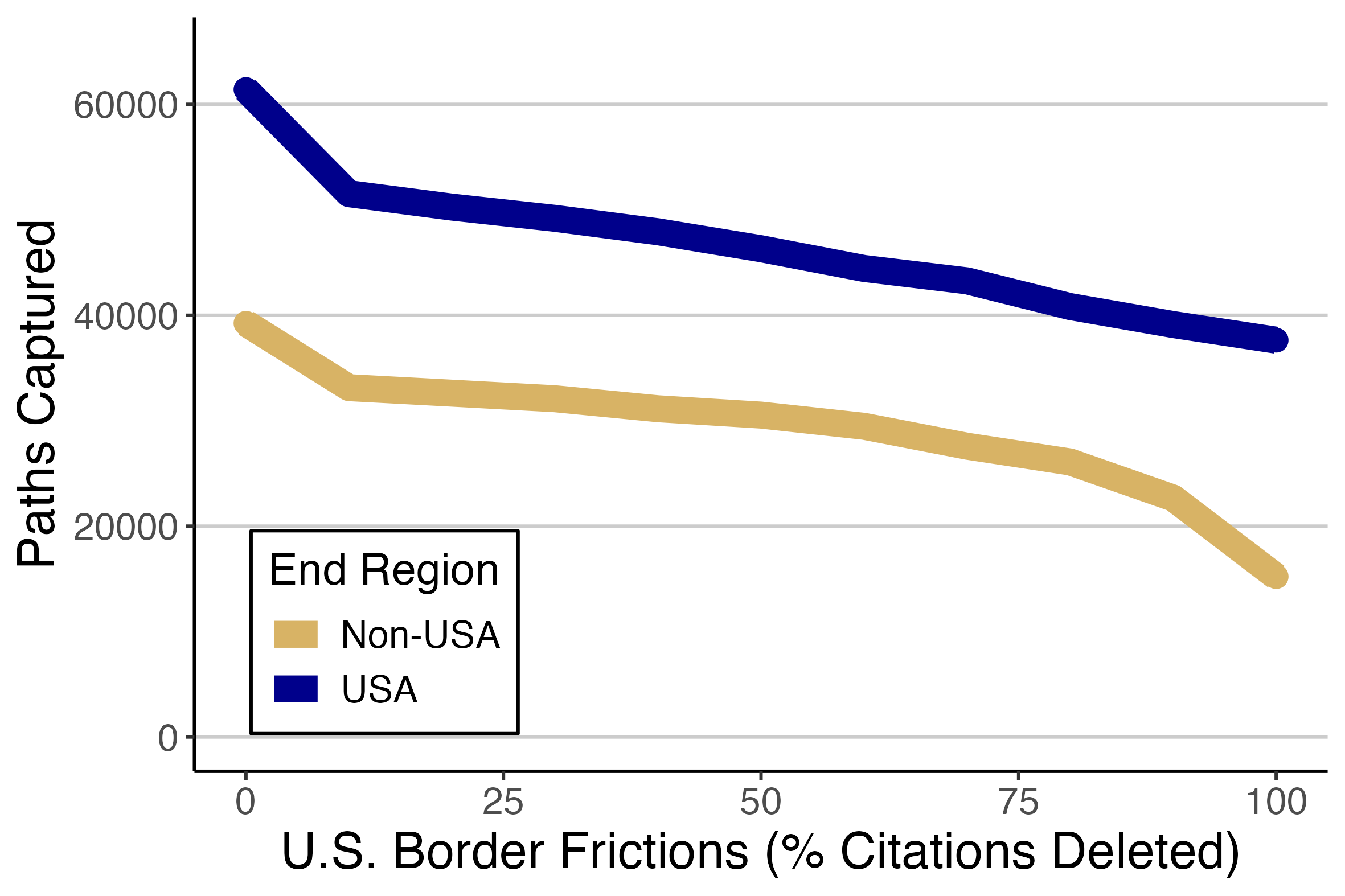}
\end{minipage}\hfill
\begin{minipage}[t]{0.48\textwidth}
\centering
\textbf{B}\\
\includegraphics[width=\textwidth]{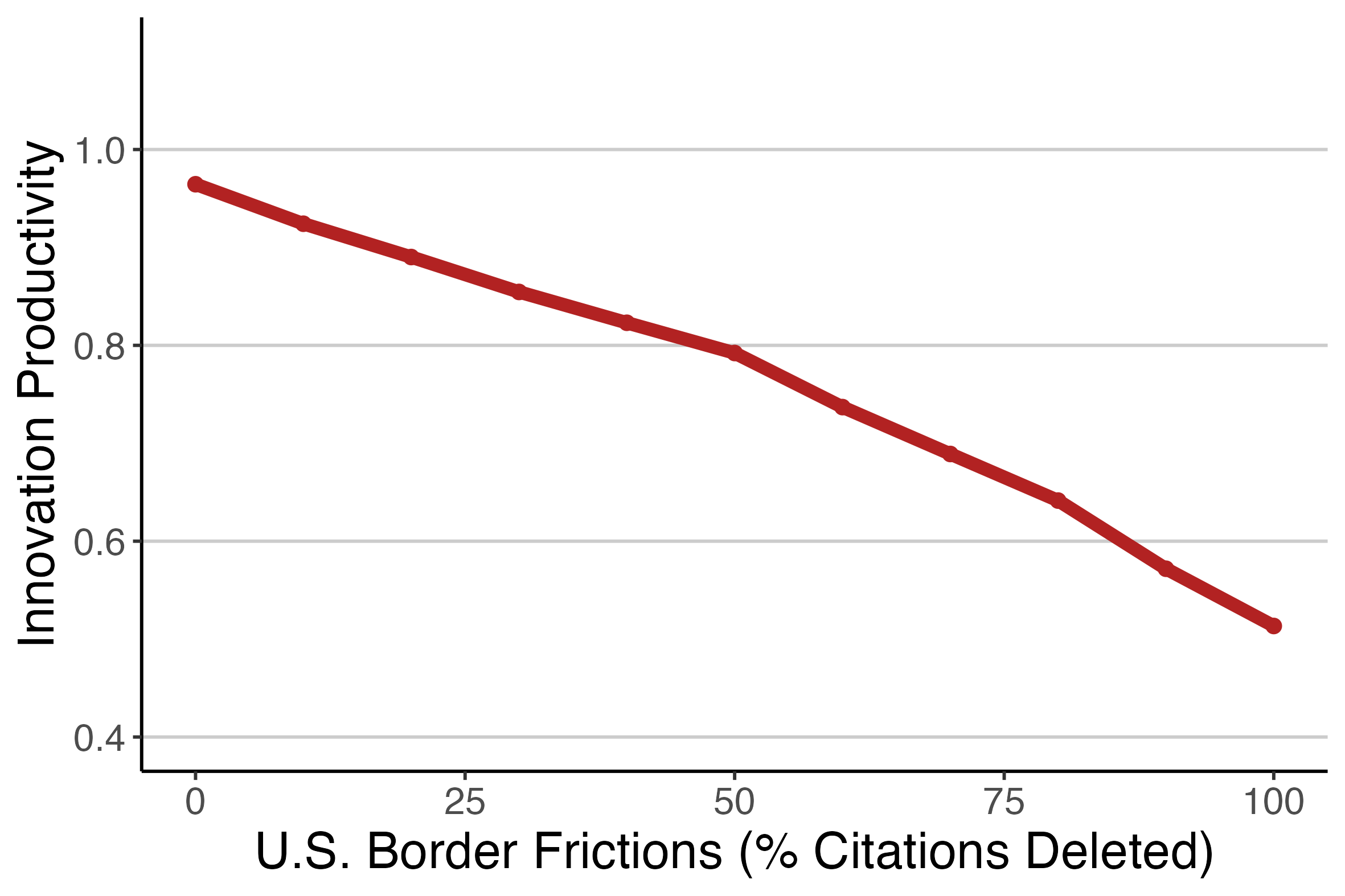}
\end{minipage}
\caption{\textbf{Short-Run Effects of Knowledge Flow Restrictions, Raw Values.} (\textbf{A}) Paths captured. (\textbf{B}) U.S. innovation productivity.}
\label{fig:s_capture_absolute_shortrun}
\end{figure}

While the main text shows the impacts of border restrictions on long-run capture and U.S. innovation productivity relative to the status quo, the following figures show capture and U.S. productivity in absolute terms. Figure~\ref{fig:s_capture_absolute_longrun} shows raw (non-normalized) values for long-run effects, while Figure~\ref{fig:s_capture_absolute_shortrun} shows raw values for short-run effects.

\subsubsection*{Border Crossings by End Region}

\begin{figure}
\centering
\includegraphics[width=0.5\textwidth]{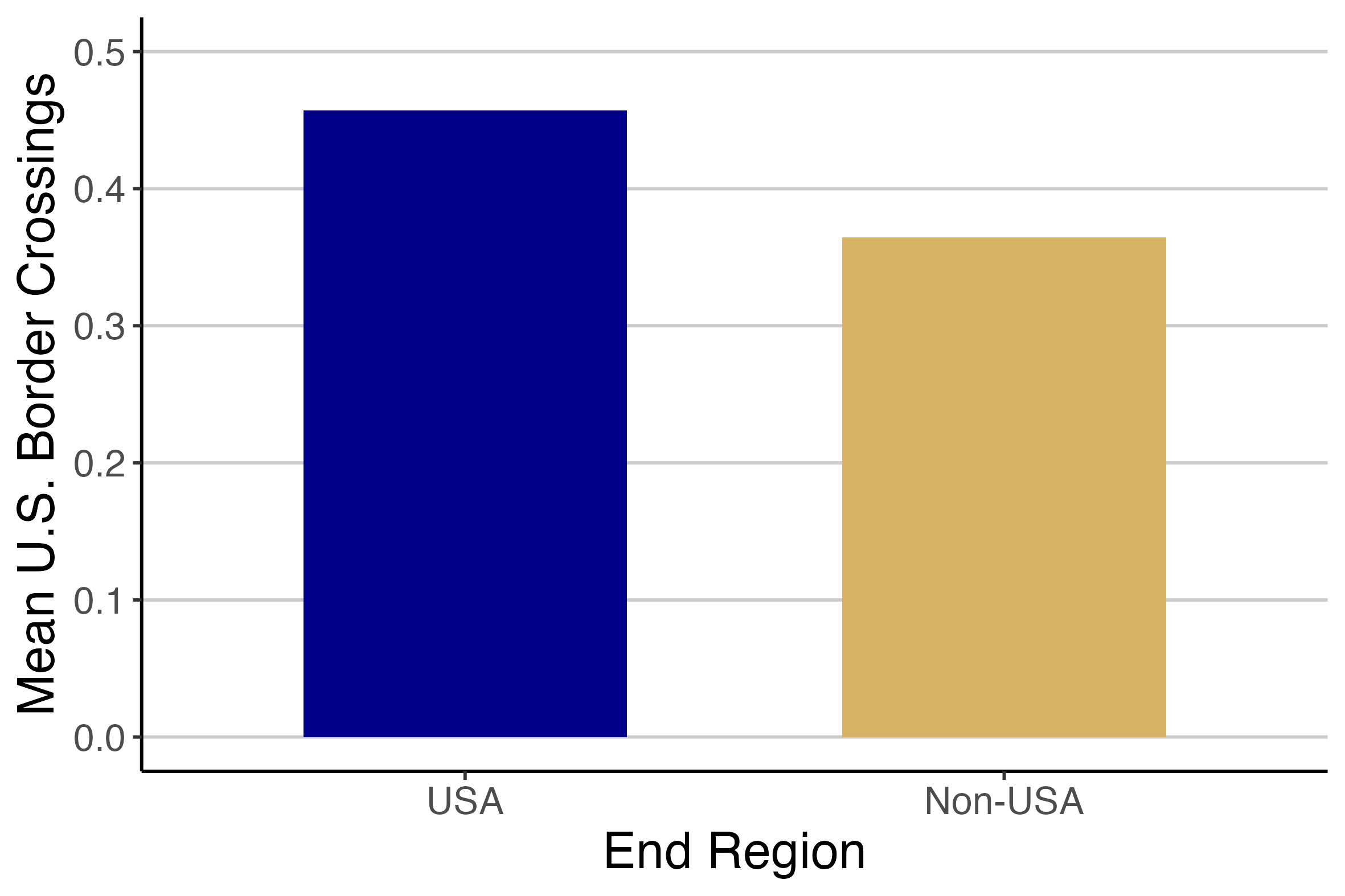}
\caption{\textbf{Mean Number of Crossings of the U.S. Border for Paths with U.S. and Non-U.S. End Regions.}}
\label{fig:s_border_crossings}
\end{figure}

\begin{figure}
\centering
\includegraphics[width=0.6\textwidth]{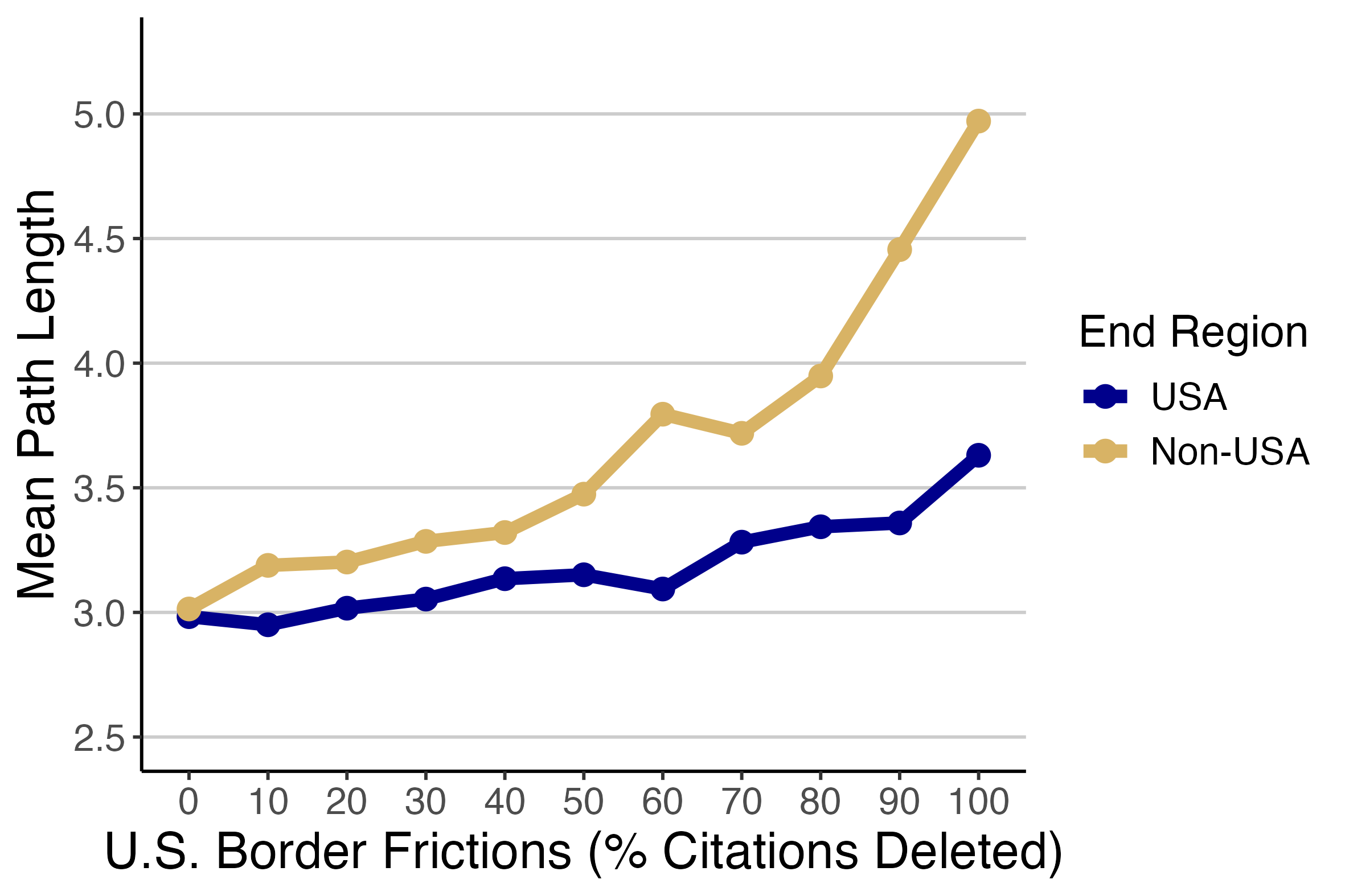}
\caption{\textbf{Path Length for U.S. and Non-U.S. Destined Paths.}}
\label{fig:s_path_length}
\end{figure}

In Figure~\ref{fig:s_border_crossings}, I show the mean number of times paths cross the U.S. border between their start point and end patent, broken out by their end region (U.S. or non-U.S.). Paths that end in the U.S. cross the border an average of 0.48 times, while those that end at non-U.S. patents cross the border an average of 0.40 times.

\subsubsection*{Change in Path Length}

Path lengths increase both inside and outside the U.S. as border frictions increase. Under the status quo, paths terminating at both U.S. and non-U.S. patents are on average 3 citations long. Under autarky, paths terminating at U.S. patents increase to an average length of 3.63, while those terminating at non-U.S. patents increase to a length of 4.97 citation steps.

\begin{figure}
\centering
\begin{minipage}[t]{0.48\textwidth}
\centering
\textbf{A}\\
\includegraphics[width=\textwidth]{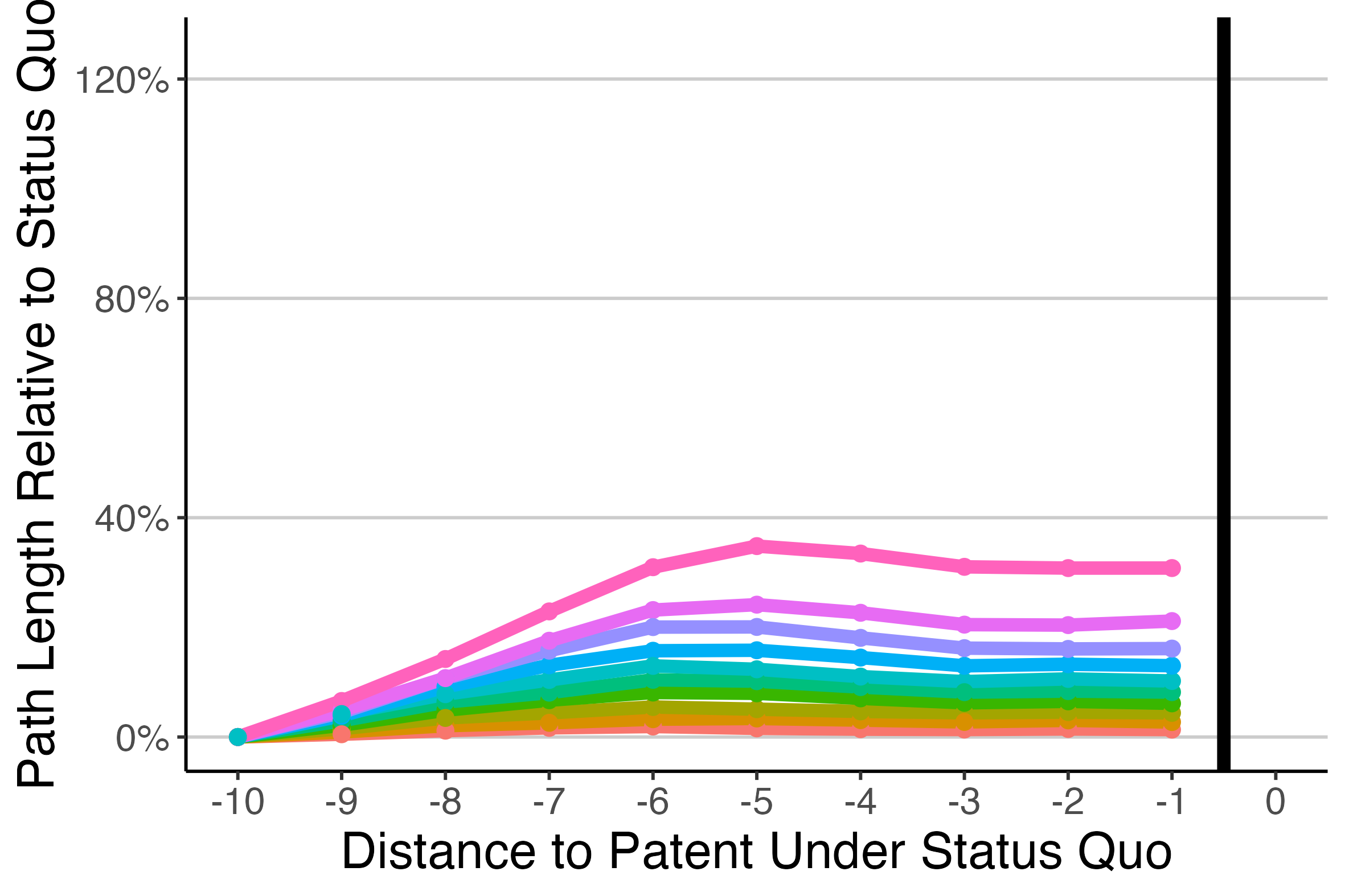}
\end{minipage}\hfill
\begin{minipage}[t]{0.48\textwidth}
\centering
\textbf{B}\\
\includegraphics[width=\textwidth]{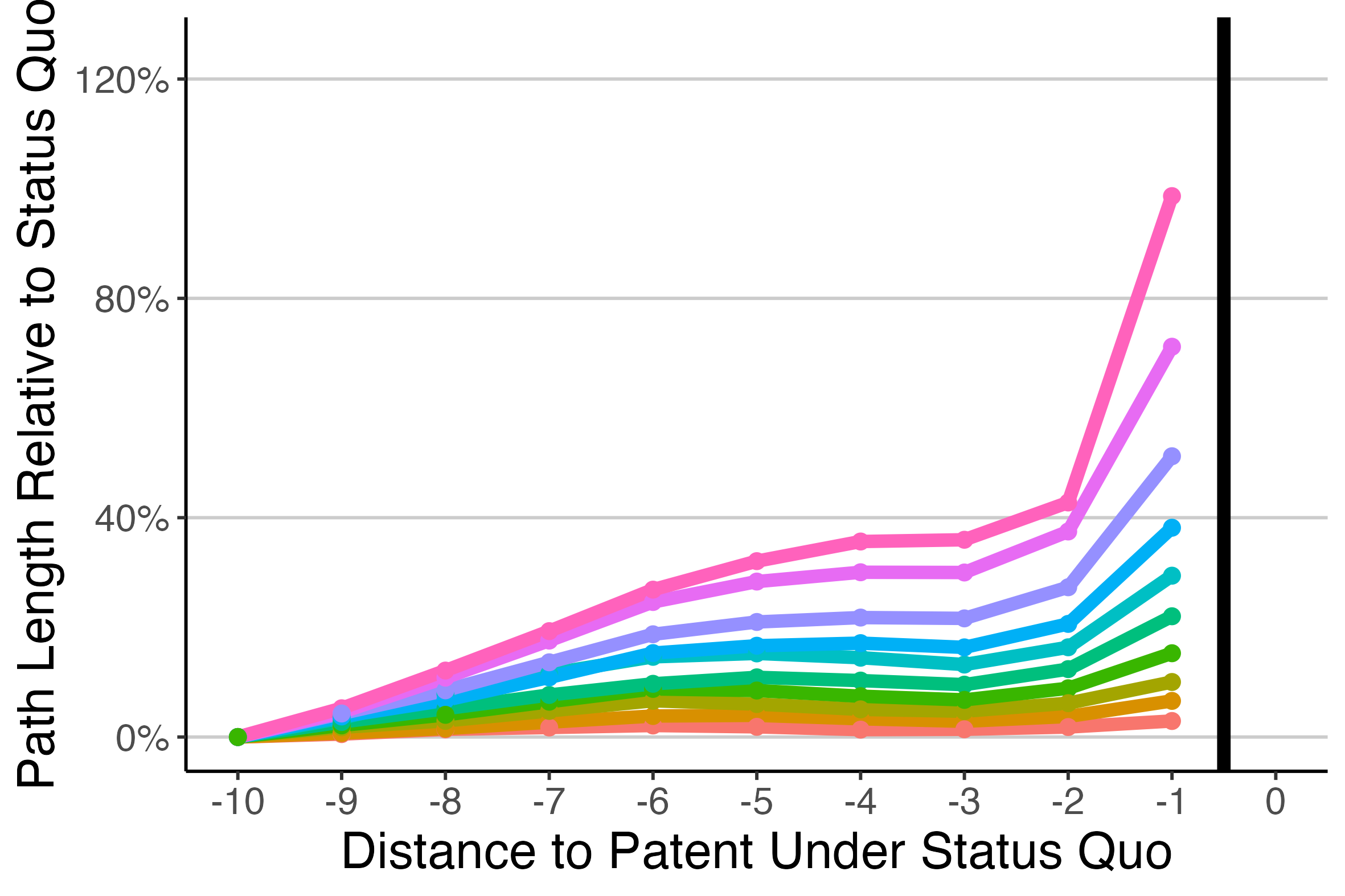}
\end{minipage}

\vspace{1em}

\includegraphics[width=0.8\textwidth]{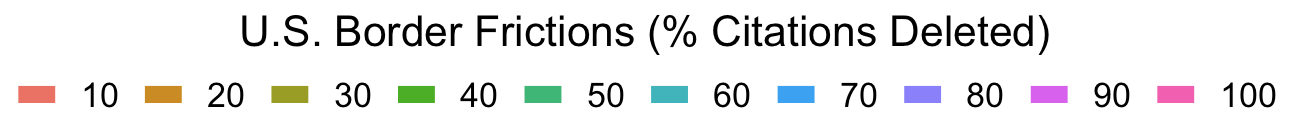}
\caption{\textbf{Change in Mean Length of Paths Connecting NSF-Supported Articles to Patents by Starting Distance.} (\textbf{A}) Paths captured by the U.S. (\textbf{B}) Paths captured outside the U.S.}
\label{fig:s_path_length_by_d}
\end{figure}

In Figure~\ref{fig:s_path_length_by_d}, I decompose these mean effects by the starting Distance of the paths. This decomposition allows one to determine whether the larger increase in mean path length for non-U.S. endpoints is driven by paths that begin close to the patent boundary, or further from the boundary. Changes in the mean path length for the various knowledge flow restriction scenarios are shown for paths ending in the U.S. in Figure~\ref{fig:s_path_length_by_d}A, and for paths ending outside the U.S. in Figure~\ref{fig:s_path_length_by_d}B.

Figure~\ref{fig:s_path_length_by_d}A shows that U.S.-captured paths increase in length by about 35\% between the status quo scenario and the U.S. autarky scenario for all StartingD values between $-6$ and $-1$. In contrast, Figure~\ref{fig:s_path_length_by_d}B shows that paths captured outside the U.S. increase by progressively larger amounts for paths that start closer to the patent boundary, when contrasting between the status quo and U.S. autarky cases. In particular, paths with $StartingD = -1$ increase in length by 98\% if they are captured outside the U.S. The much larger increase for paths that start near the patent boundary and are captured outside the U.S. suggests that paths captured outside the U.S. struggle to form new connections to patents when knowledge flows across the U.S. border are impeded.

\subsubsection*{Change in U.S. Share of Intermediary Science}

\begin{figure}
\centering
\includegraphics[width=0.6\textwidth]{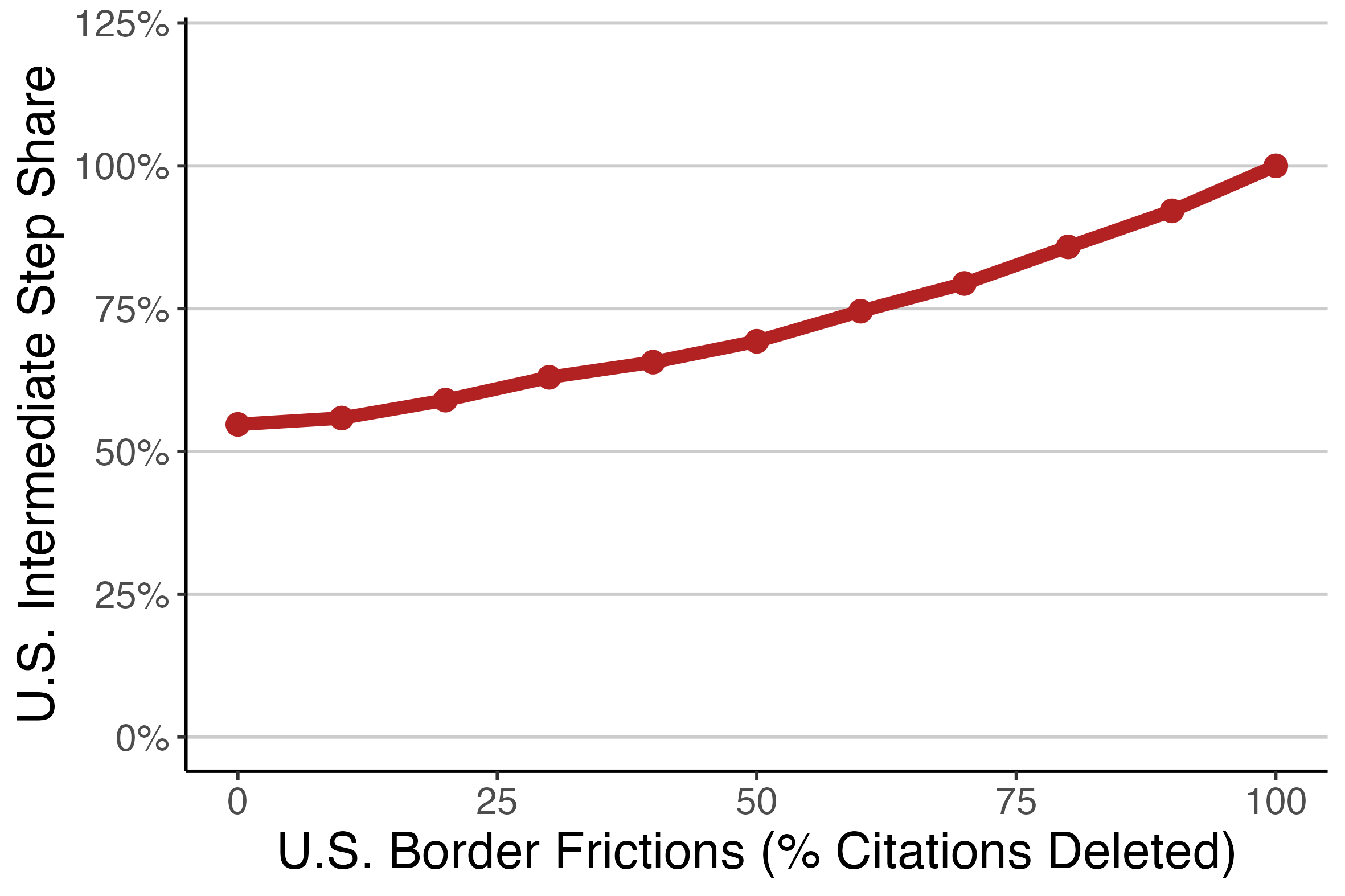}
\caption{\textbf{Share of Intermediary Papers Linked to U.S. Patents that are Produced in the U.S.}}
\label{fig:s_intermediary_steps_US}
\end{figure}

As border restrictions increase, more of the intermediary science linked to U.S. patents is conducted in the U.S. As Figure~\ref{fig:s_intermediary_steps_US} shows, under the status quo, 54.7\% of U.S.-linked intermediary papers are produced in the U.S. However, under autarky that share rises to 100\%.

\subsubsection*{Path Capture by U.S. Metropolitan Area}

\begin{figure}
\centering
\includegraphics[width=\textwidth]{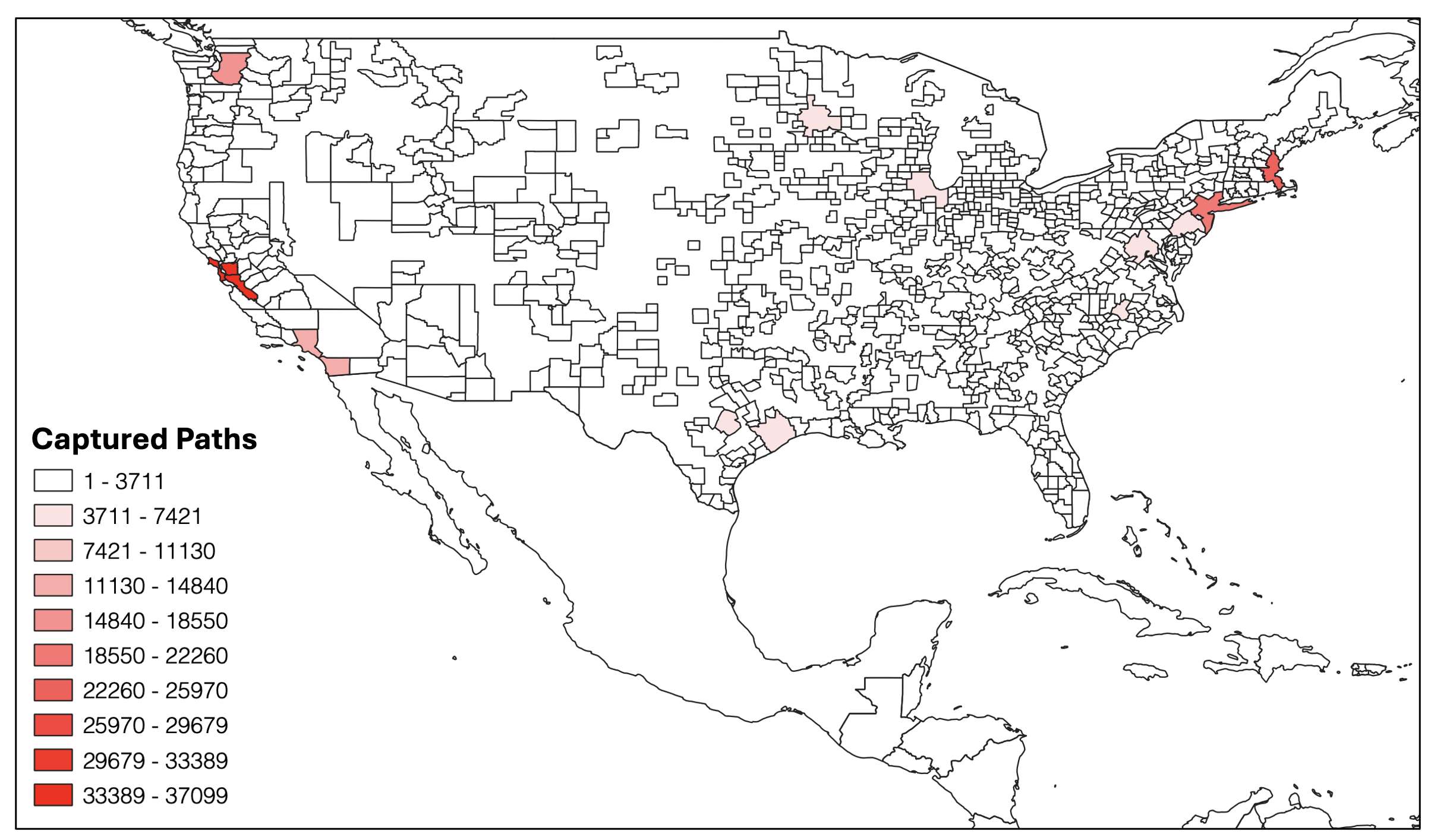}
\caption{\textbf{Number of Paths Captured by U.S. Metropolitan Area.}}
\label{fig:s_map_metros}
\end{figure}

Figure~\ref{fig:s_map_metros} presents a map of U.S. metropolitan areas by their number of captured paths. There are very strong concentrations in the metropolitan areas centered on San Francisco and San Jose, and weaker concentrations in the metro areas centered on New York City, Boston, Seattle, Los Angeles, and San Diego. All other metro areas capture comparatively few paths. Moreover, the geography of path capture is similar to the geography of venture capital, suggesting that the United States is able to translate globally-sourced science into patents because it has agglomerations that efficiently match scientific knowledge to financial resources \cite{samila_venture_2010,samila_venture_2011}.

\subsubsection*{Short-Run Path Capture by Firms}

\begin{figure}
\centering
\includegraphics[width=\textwidth]{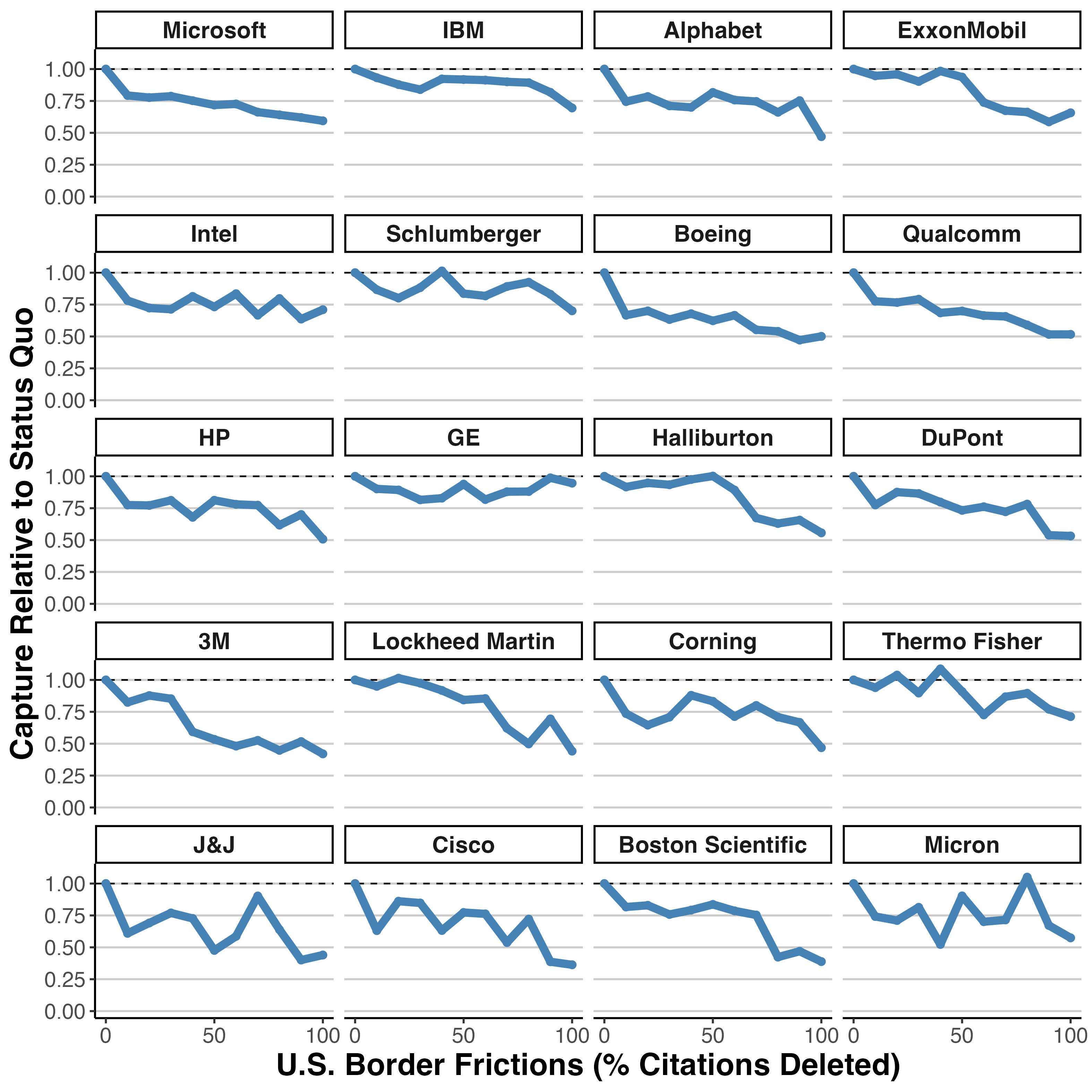}
\caption{\textbf{Change in Outstanding Path Capture at the Top 20 U.S. Corporations Reliant on NSF-Stimulated Knowledge Paths.}}
\label{fig:s_firms_path_capture}
\end{figure}

Figure~\ref{fig:s_firms_path_capture} shows how knowledge flow barriers affect outstanding path capture at U.S. corporations. The 20 corporations with the most patents linked to NSF-funded science are shown. Knowledge flow restrictions reduce path capture at 19 of these corporations, with sizable impacts at companies like Microsoft, Alphabet (Google), ExxonMobil, Intel, Boeing, Qualcomm, Hewlett Packard (HP), Halliburton, DuPont, 3M, Lockheed Martin, Corning, Johnson and Johnson, Cisco, and Boston Scientific. These results indicate that restrictions impeding the flow of science across the U.S. border would reduce innovative capture at many of the U.S.' major corporations.

\subsubsection*{Impacts at the Country and Regional Level}

In this section, I replicate the main analyses while disaggregating non-U.S. captured paths into their respective regions.

\begin{figure}
\centering
\includegraphics[width=0.6\textwidth]{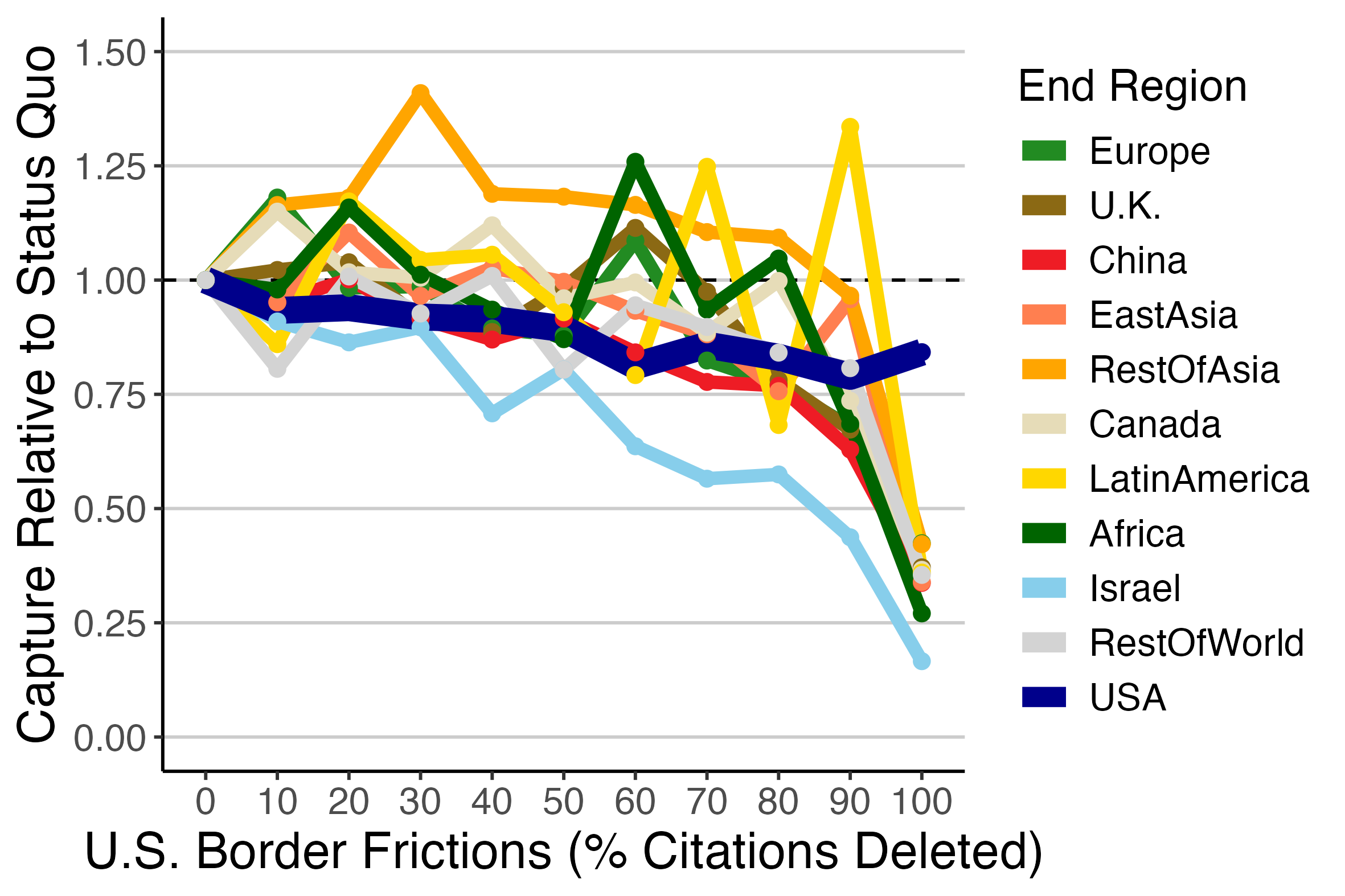}
\caption{\textbf{Long-Run Change in Number of Realized Paths Captured, by All Regions.}}
\label{fig:s_allregions_capture_longrun}
\end{figure}

Figure~\ref{fig:s_allregions_capture_longrun} shows the change in the number of realized paths captured by each region, by knowledge flow barrier levels. Notably, knowledge barriers at the U.S. border reduce the number of paths captured by Israel, but increase the number captured in the Rest of Asia and Canada (except in the cases with very high knowledge flow barriers). The number of paths captured by Africa increases in percentage terms, but the raw number captured in Africa is very small, so this result is not meaningful.

\begin{figure}
\centering
\includegraphics[width=0.6\textwidth]{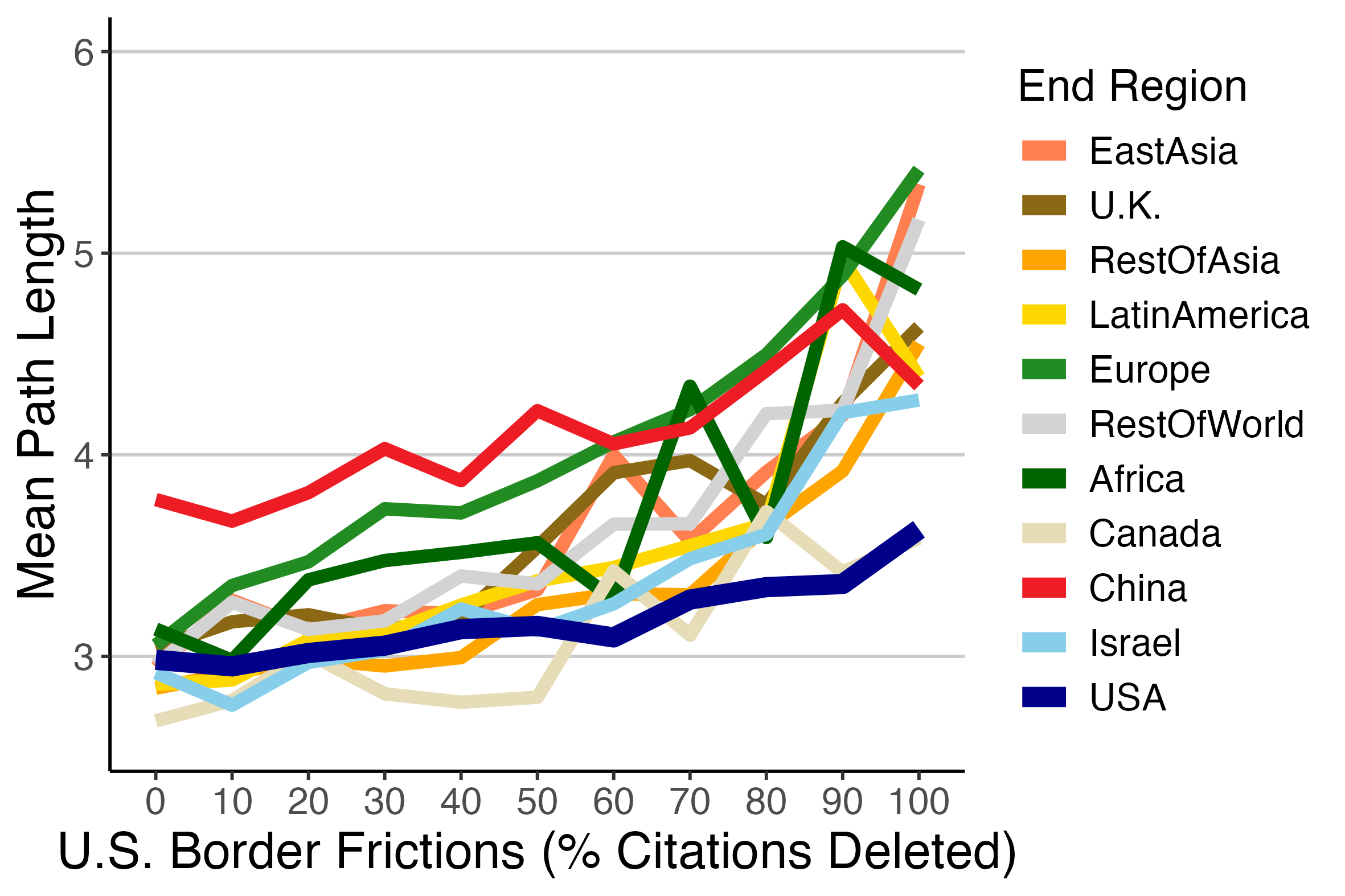}
\caption{\textbf{Long-Run Mean Length of Realized Paths, for All Regions.}}
\label{fig:s_allregions_pathlength}
\end{figure}

Figure~\ref{fig:s_allregions_pathlength} shows the mean length of paths by region of capture. Mean path length under the status quo is notably higher in Israel and the Rest of Asia. Mean path length increases for all regions as knowledge flow barriers at the U.S. border increase; however, paths captured by the U.S. increase in length less than do those that end in the other regions.

\begin{figure}
\centering
\includegraphics[width=0.6\textwidth]{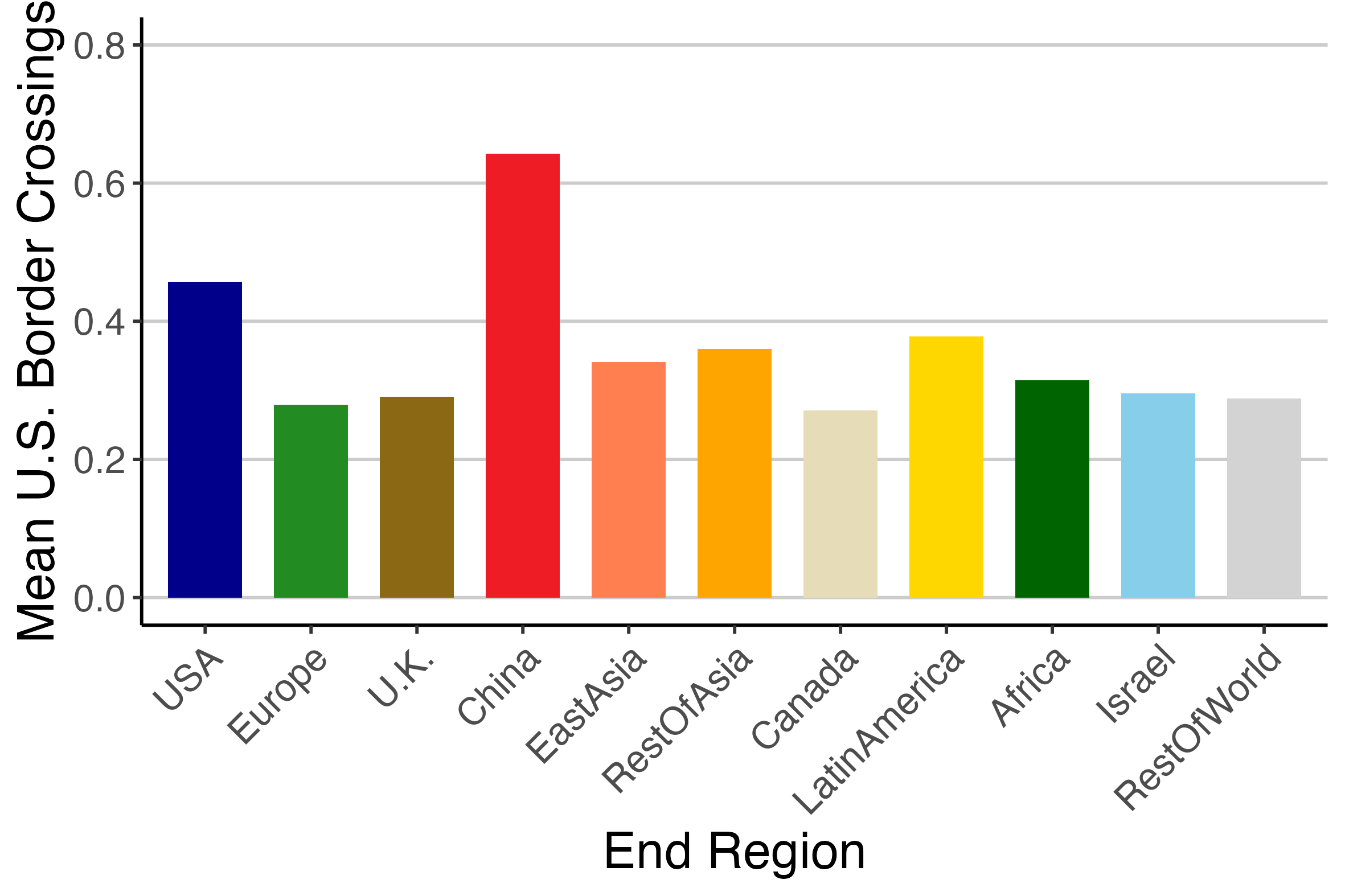}
\caption{\textbf{Number of Path Crossings of the U.S. Border by End Region.}}
\label{fig:s_allregions_border_crossings}
\end{figure}

Figure~\ref{fig:s_allregions_border_crossings} shows the mean number of crossings of the U.S. border by capturing region. Paths captured by Israel cross the U.S. border the most times, suggesting that these paths are most vulnerable to knowledge flow restrictions at the U.S. border. Notably, paths captured by the U.S. cross the U.S. border more times than those captured by any other region, with the exception of Israel.

\begin{figure}
\centering
\begin{minipage}[t]{0.48\textwidth}
\centering
\textbf{A}\\
\includegraphics[width=\textwidth]{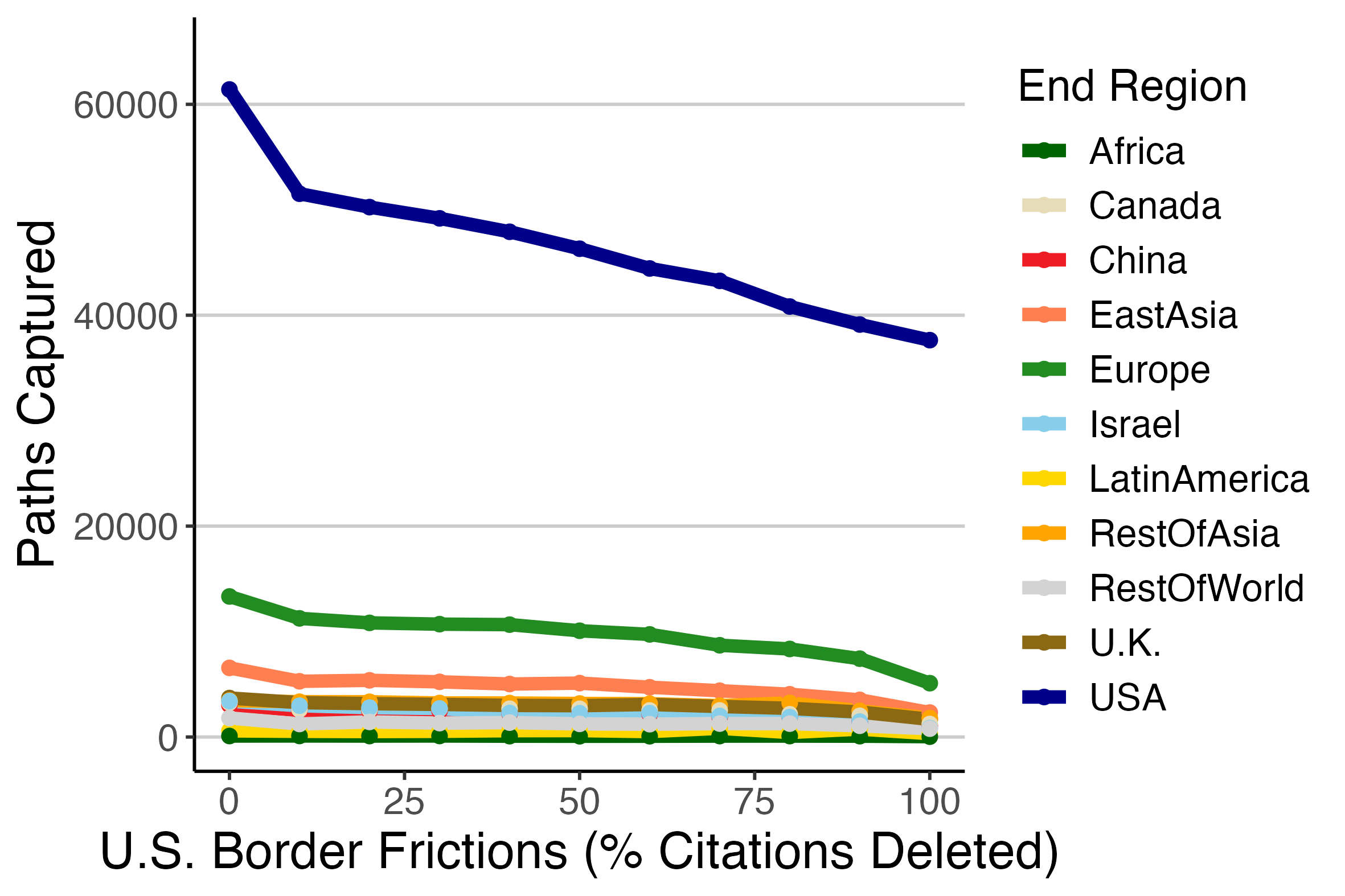}
\end{minipage}\hfill
\begin{minipage}[t]{0.48\textwidth}
\centering
\textbf{B}\\
\includegraphics[width=\textwidth]{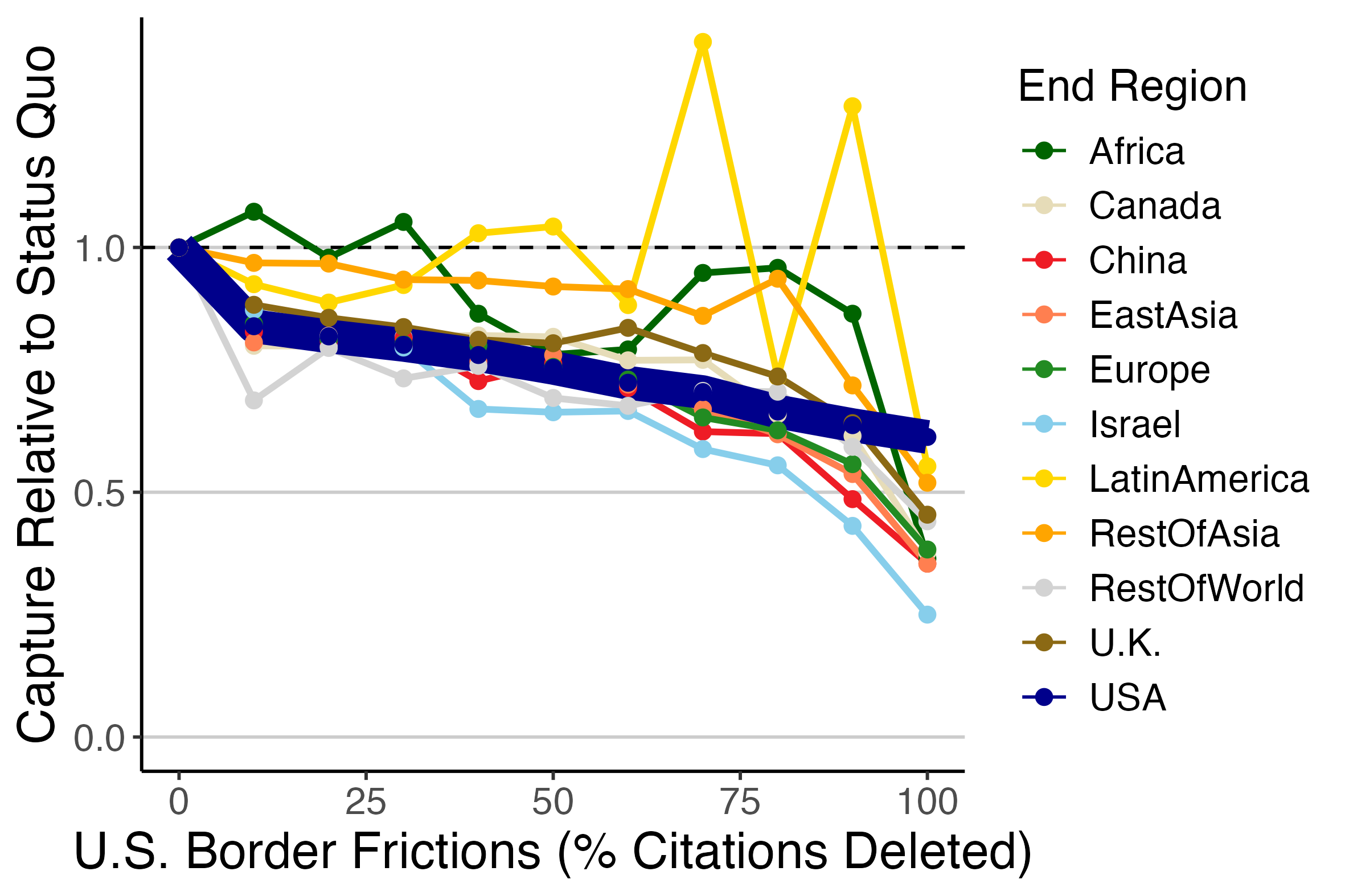}
\end{minipage}
\caption{\textbf{Short-Run Paths Captured by Region.} (\textbf{A}) Number of paths captured by region. (\textbf{B}) Change in paths captured by region.}
\label{fig:s_allregions_capture_shortrun}
\end{figure}

Figure~\ref{fig:s_allregions_capture_shortrun} shows information on the short-run impacts on capture by region. Figure~\ref{fig:s_allregions_capture_shortrun}A shows the number of captured outstanding paths, while Figure~\ref{fig:s_allregions_capture_shortrun}B shows the change in the number of outstanding paths.

\end{document}